# A Map of the Inorganic Ternary Metal Nitrides

Wenhao Sun[1], Christopher Bartel[2], Elisabetta Arca[3], Sage Bauers[3], Bethany Matthews[4], Bernardo Orvañanos[5], Bor-Rong Chen,[6] Michael F. Toney,[6] Laura T. Schelhas,[6] William Tumas[3], Janet Tate,[4] Andriy Zakutayev[3], Stephan Lany[3], Aaron Holder[2,3], Gerbrand Ceder[1,7]

[1] Materials Sciences Division, Lawrence Berkeley National Laboratory, Berkeley, California 94720, USA

[2] Department of Chemical and Biological Engineering, University of Colorado, Boulder, Colorado 80309, USA

[3] National Renewable Energy Laboratory, Golden, Colorado 80401, USA

[4] Department of Physics, Oregon State University, Corvallis, Oregon 97331, USA

[5] Department of Materials Science and Engineering, Massachusetts Institute of Technology, Cambridge, MA 02139

[6] SLAC National Accelerator Laboratory, Menlo Park, CA, 94025, USA.

[7] Department of Materials Science and Engineering, UC Berkeley, Berkeley, California 94720, USA

Corresponding Authors: wenhaosun@lbl.gov, aaron.holder@colorado.edu

**Abstract:**

Exploratory synthesis in novel chemical spaces is the essence of solid-state chemistry. However, uncharted chemical spaces can be difficult to navigate, especially when materials synthesis is challenging. Nitrides represent one such space, where stringent synthesis constraints have limited the exploration of this important class of functional materials. Here, we employ a suite of computational materials discovery and informatics tools to construct a large stability map of the inorganic ternary metal nitrides. Our map clusters the ternary nitrides into chemical families with distinct stability and metastability, and highlights hundreds of promising new ternary nitride spaces for experimental investigation—from which we experimentally realized 7 new Zn- and Mg-based ternary nitrides. By extracting the mixed metallicity, ionicity, and covalency of solid-state bonding from the DFT-computed electron density, we reveal the complex interplay between chemistry, composition, and electronic structure in governing large-scale stability trends in ternary nitride materials.

Nitrides are an exciting space for materials design,[1,2,3] as exemplified by state-of-the-art nitride materials for solid-state lighting,[4,5] ceramic hard coatings,[6] ammonia-synthesis catalysts,[7,8] permanent magnets,[9] superconductors,[10] superinsulators,[11] electrides[12] and more. The nitride ($N^{3-}$) anion imparts unique electronic and bonding characteristics that are difficult to achieve in other chemical spaces, including hybridization of nitrogen 2*p* states with metal *d* states for useful optoelectronic and defect-tolerance properties,[13] as well as the formation of strong metal-nitrogen bonds leading to structural stability and mechanical stiffness.[14] Despite much promise in the functionality of nitride materials, the nitrides are relatively underexplored, with fewer than 400 unique ternary metal nitrides catalogued in the Inorganic Crystal Structure Database (ICSD) in contrast to over 4,000 ternary metal oxides. The paucity of known nitrides can largely be attributed to the challenging requirements of nitride synthesis. Because the $N_2$ molecule is so stable, solid-state nitrides generally have small formation energies; they decompose at high-temperature; and they must be synthesized in oxygen- and water-free atmospheres to achieve high purity.[2,15,16,17] These stringent synthesis constraints, coupled with the poor intrinsic stabilities of nitrides, impose significant risk on the exploratory synthesis of novel nitride materials.

High-throughput computational materials science has emerged as a new paradigm for materials discovery,[18,19] helping to guide experimental synthesis efforts across broad and uncharted chemical spaces. Here, we employ a suite of computational materials discovery[20,21,22,23] and informatics[24,25] tools to survey, visualize, and most importantly, explain stability relationships across the inorganic ternary metal nitrides. Our investigation proceeds in three steps. First, we use crystal structure prediction algorithms to probe the energy landscapes of previously unexplored ternary nitride spaces, surveying novel nitride compounds over 962 $M_1$-$M_2$-N spaces. We predict a large list of new stable and metastable ternary nitrides, significantly extending the known thermochemical data in this space. Guided by these predictions, we experimentally synthesize 7 new Zn- and Mg-based ternary nitrides, and identify hundreds of promising new ternary nitride systems for further exploratory synthesis.

Much like how Mendeleev's Periodic Table revealed the underlying structure of the elements, an effective visual organization can reveal hidden relationships and chemical families within the ternary metal nitrides. Assisted by unsupervised machine-learning algorithms,[26] we next cluster together metals that have a similar propensity to form stable or metastable ternary nitrides. We use these clustered nitride families to construct a large and comprehensive stability map of the inorganic ternary metal nitrides. Not only does our map visualize broad overarching relationships between nitride chemistry and thermodynamic stability, it further inspires us to rationalize these trends from their underlying chemical origins.[27,28] To do so, we extract from the DFT-computed electron density the mixed metallicity, ionicity, and covalency of solid-state bonding—providing new chemical features to interpret the electronic origins of nitride stability. We show that the nitrogen anion can be surprisingly amphoteric in the solid-state, usually acting as an electron acceptor in nitrogen-rich nitride ceramics, but remarkably, sometimes serving as an electron donor to stabilize nitrogen-poor metallic nitrides.

Beyond the nitrides, there remain many other unexplored chemical spaces awaiting experimental discovery. Our computational approach here can be further applied to these uncharted chemical spaces, not only to predict and synthesize new compounds, but also to visualize general trends over broad compositional spaces—providing maps and chemical intuition to help experimental chemists more rationally navigate exploratory synthesis at the frontier of solid-state chemistry.

## Results

In this work, we explore ternary nitrides over a 50×50 $M_1$-$M_2$-N composition space, where M consists of the 50 most common cations in the known nitrides. These cations broadly sample the periodic table; spanning over the alkali, alkaline earth, transition, precious, and post-transition metals, as well as the main group elements B, C, Si, S and Se. Within this composition matrix, known ternary nitride compounds exist over only 303 $M_1$-$M_2$-N spaces. To fill in the missing spaces, we first conduct a high-throughput computational search for novel ternary nitride compounds. Previous computational searches for ternary nitrides have been constrained to either limited composition spaces[29,30,31] or specific crystal structures.[32,33,34] Here, we broadly sample over both composition and crystal structure, using a data-mined structure predictor (DMSP)[35] to perform rational chemical substitutions on the known ternary nitrides, creating unobserved but reasonable novel ternary nitride phases *in silico*.

In a previous work, we trained a DMSP specifically for nitrides discovery, by data-mining which chemical substitutions in the solid-state pnictides are statistically probable.[36] We found a substitution matrix trained on a pnictides training-set to be more predictive for nitride discovery than a substitution matrix trained over all inorganic solids, which otherwise becomes biased towards ionic substitutions that are common in the more thoroughly-explored oxides and chalcogenides. In general, the substitution relationships in oxides are not applicable to nitrides due to differences in structure-types, elemental coordination, and metal redox chemistry for $O^{2-}$ vs. $N^{3-}$ anions. Here with the pnictides-trained DMSP, we extrapolate the 340 known ternary nitrides (213 stable + 127 metastable) to 6,000 hypothetical ternary nitride structures, sampled over 962 ternary $M_1$-$M_2$-N spaces.

Using density functional theory (DFT), we compute the formation enthalpies of these DMSP-generated nitrides, which are then used to probe the stability landscapes of unexplored ternary nitride spaces. We evaluate the phase stability of these candidate structures leveraging the tools and precomputed data from the Materials Project database.[37,38] **Table 1** summarizes the results from this screening. Notably, we predict 203 new stable ternary nitride compounds, nearly doubling the 213 previously-known stable ternary nitrides. These stable ternary nitrides span 277 ternary $M_1$-$M_2$-N spaces, 92 of which were not previously known to contain any stable ternary compounds. We have made the structures and energies of the newly predicted nitrides freely available on the Materials Project for readers interested in further investigation. We note that a small subset of these stable ternary nitrides have been identified in previous computational searches,[29,30,31,32,33,34] which we have reconfirmed here.

Permuting chemistry and crystal structure on the known ternary nitrides offers a computationally efficient probe of formation energies over broad ternary nitride compositions. One limitation of the DMSP is that if the structural prototype of a ground-state nitride has never been observed before, then the DMSP cannot predict it. Nevertheless, because most ternary nitride spaces are unexplored, the prediction of any ternary nitride structure with negative formation energy in an otherwise empty chemical space implies that the true ground-state structures and compositions must be even lower in energy—therefore highlighting that ternary space as a compelling target for further theoretical and experimental investigations.

A large list of predicted compounds is difficult to navigate and does not provide an intuitive picture of the structural form of a chemical space.[39] When Mendeleev constructed the Periodic Table, he produced a conceptual framework to orient our understanding of the relationships and trends between the chemical elements. In higher-order chemical spaces—binaries, ternaries, etc.—these trends become increasingly challenging to extract by hand. Here, we elucidate the structural form of the ternary nitride space using hierarchical agglomeration,[40] which is an unsupervised machine-learning algorithm, to cluster together metals with a similar propensity to form either stable or metastable ternary nitrides. In order to capture both large-scale stability trends, as well as local chemical relationships, we build a multi-feature distance metric that considers for each ternary nitride whether it is stable or metastable, its formation energy, and which periodic group the metal lies in. These multiple features represent mixed data-types (nominal, continuous and ordinal, respectively), which we combine into a single distance metric using Gower's method (details in **S.I.3**).[41]

The agglomeration algorithm clusters elements hierarchically by minimizing this multi-feature distance matrix. The resulting dendrogram provides a phenotypic representation of the nitride chemical families, grouped by their thermochemical stabilities. With this 1D ordering of metals, we produce a clustered heat map of the inorganic ternary metal nitrides, shown in **Figure 1**, colored by the stability of the lowest formation-energy ternary metal nitride in each $M_1$-$M_2$-N chemical space. Our clustering algorithm parses the ternary nitrides map into distinct regions of stability (blue), metastability against binaries (green), and metastability against elements (red)—highlighting stable ternary nitride spaces that are promising for further exploratory synthesis, and metastable spaces where successful synthesis may require non-equilibrium synthesis routes. An interactive version of the map, with ternary phase diagrams and compound stability information for each $M_1$-$M_2$-N system, is available in **S.I.1**. (*For this preprint, an interactive version is at* https://wenhaosun.github.io/TernaryNitridesMap.html)

**Table 1.** Statistics of the known and predicted ternary metal nitrides, categorized by the thermodynamic stability of ternary $M_1$-$M_2$-N spaces, and specific ternary $A_xB_yN_z$ phases within those spaces. All spaces are categorized by the $\Delta H_f$ of the lowest formation-energy ternary nitride. Metastable phases are categorized by their energy above the convex hull, $\Delta E_{hull}$.

| Ternary $M_1$-$M_2$-N Spaces | Previously Known | Newly Predicted | Portion of map |
|---|---|---|---|
| Systems with Stable Ternary Nitrides (Blue) | 189 | 92 | 281 (29%) |
| - Stable Alkali-Metal-Nitride Systems | *124* | *76* | *200* |
| - Stable Metal-Metal-Nitride Systems | *65* | *16* | *81* |
| Metastable vs. Stable Binaries, $\Delta H_f < 0$ (Green) | 98 | 322 | 420 (44%) |
| Metastable vs. Elements, $\Delta H_f > 0$ (Red) | 20 | 241 | 261 (27%) |

| Ternary $A_xB_yN_z$ Phases | Previously Known | Newly Predicted | Total Number |
|---|---|---|---|
| Stable Ternary Phases | 213 | 203 | 416 |
| Metastable, $\Delta E_{Hull} < 70$ meV/atom | 39 | 36 | 75 |
| Metastable, $\Delta E_{Hull} < 200$ meV/atom | 85 | 175 | 260 |
| Metastable: Stabilizable $\Delta\mu N < +1$ eV/N | 3 | 92 | 95 |

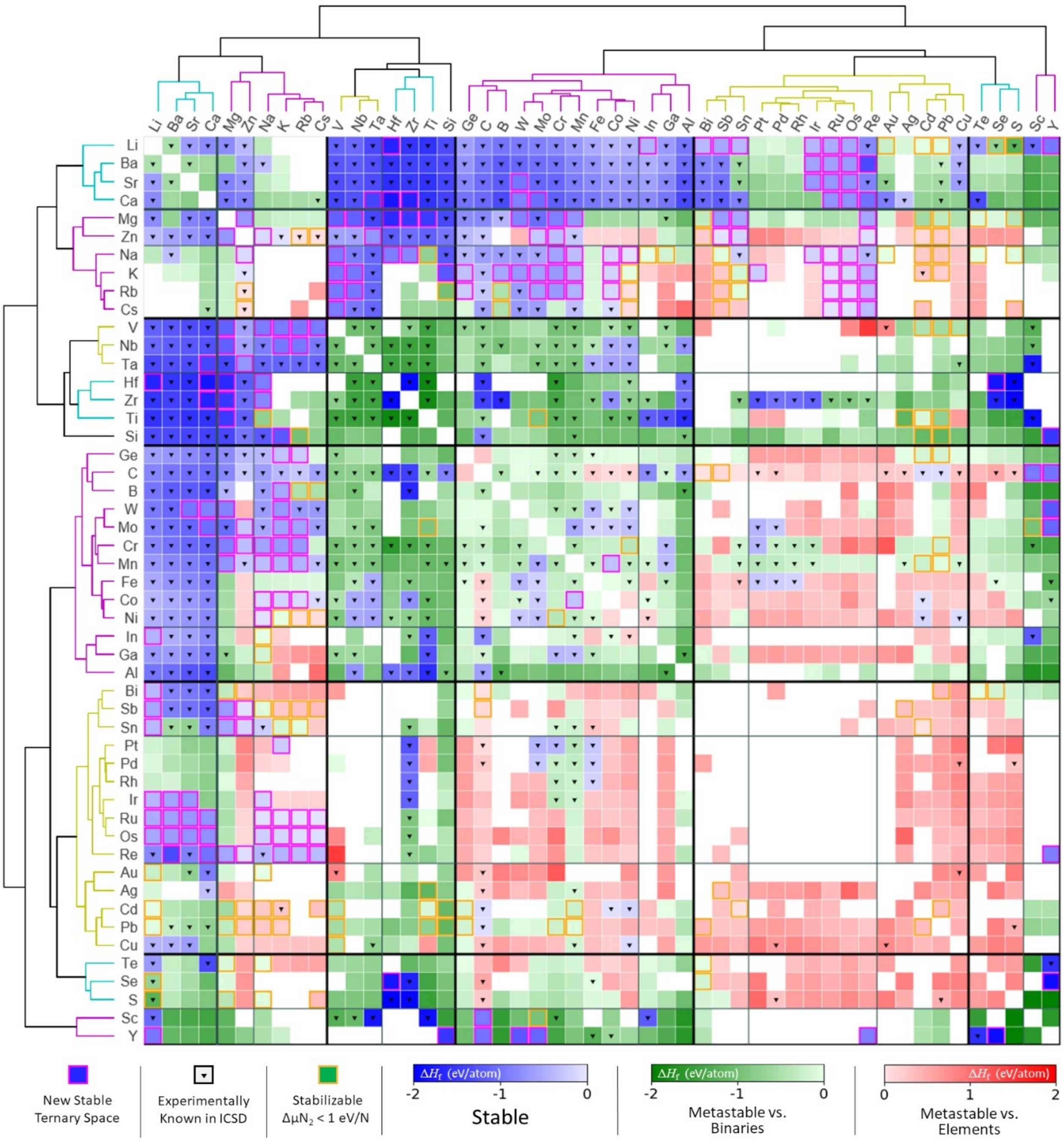

**Figure 1.** Map of the inorganic ternary metal nitrides, colored by the thermodynamic stability of the lowest formation-energy ternary nitride. Blue: stable ternary nitrides on the convex hull; Green: ternaries with $\Delta H_f < 0$ but metastable with respect to binaries; Red: ternaries metastable with respect to elements, $\Delta H_f > 0$. Triangles represent ternary nitride systems with entries in the ICSD. White spaces indicate that the DMSP did not find probable chemical substitutions to create a structure in that system. Elements are clustered on multiple features to indicate their propensity to form stable or metastable ternary nitrides. These clustered elements are represented phenotypically by a dendrogram, which parses the ternary nitrides map into regions of distinct stability and metastability. An interactive version of the ternary nitride map, with phase diagrams and compound stability information for each ternary system, is available in **S.I.1**. (*For this preprint, an interactive version is at* https://wenhaosun.github.io/TernaryNitridesMap.html)

## Stable Ternary Nitrides

Thermodynamically stable ternary nitrides are relatively rare, comprising only a quarter of the map, which is likely a confounding factor in the difficulty of ternary nitride discovery. Alkali and alkaline earth ternary nitrides (Alk-Me-N) represent a majority of the stable ternary nitride spaces (200/281 = 71%), whereas stable non-alkali metal-metal-nitrides (Me-Me-N) are less common, with small islands of stability scattered amongst the mixed transition- and precious-metal nitrides.

The clustered dendrogram distinguishes between three major groups of alkali/alkaline earth elements in their ability to form ternary nitrides. The first group is composed of Li, Ca, Sr, and Ba, which form ternary nitrides with negative $\Delta H_f$ with all elements, most of which are thermodynamically stable. The alkali ions Na, K, Rb, and Cs also form stable ternaries with early and first-row transition metals, although they generally react unfavorably ($\Delta H_f > 0$) with precious metals and metalloids. The clustering algorithm places Mg and Zn as intermediate between these two groups. Although one might anticipate Mg to be chemically similar to Ca, Sr and Ba, Mg is less reactive than the other alkaline earth metals–forming ternaries less exothermically and forming fewer stable ternary nitrides overall. For a transition metal, Zn is relatively electropositive, meaning it can react like an alkali when coupled with early transition metals; but when coupled with late- and post-transition metals, Zn-containing ternaries generally have positive formation energy.

Of the 281 spaces with stable ternary nitrides, the 92 indicated on the map by a magenta box do not have any ternary nitride entries in the current ICSD, and therefore represent new theoretical predictions. While most stable ternary nitrides containing Li, Ca, Ba, and Sr have been synthesized experimentally, ternary nitrides with Na, K, Rb, and Cs have not been as readily explored. This may be because Na, K, Rb, and Cs do not form stable binary nitrides, meaning that their precursors for solid-state synthesis are less conveniently available. Nevertheless, these compositions are promising for further experimental synthesis. Scandium and yttrium form several new stable ternaries, with some of the most negative formation energies on the map. However, the large exothermic formation energies of binary ScN and YN renders most Sc- and Y-based ternary nitrides to be metastable against decomposition, to the extent that the clustering algorithm categorizes Sc and Y as independent chemistries from the rest of the transition metals. Interestingly, we predict the precious metals Ir, Ru, Re, Os to form stable ternary nitrides when synthesized with most alkali and alkaline earth metals, representing new families of stable ternary metal nitrides that await experimental discovery.

## Experimental Validation of Predicted Ternary Nitrides

From our predictions, we identified Zn- and Mg-based ternary nitrides as compelling target spaces for novel materials synthesis. Using magnetron sputtering, we successfully synthesized crystalline nitride thin-films in 7 new ternary nitride spaces: Zn-Mo-N, Zn-W-N, Zn-Sb-N, Mg-Ti-N, Mg-Zr-N, Mg-Hf-N, Mg-Nb-N. Concurrently, we conducted an unconstrained DFT search of ground-state structures and compositions using Kinetically Limited Minimization,[42] with resulting compositions, structures and formation energies tabulated in **Figure 2a**. As illustrated in **Figure 2b**, these Zn-based ternaries adopt a wurtzite-derived structure, whereas the Mg-based ternaries form in a rocksalt-derived structure. **Figure 2c** shows experimental synchrotron X-ray diffraction (XRD) patterns of these novel synthesized nitrides together with reference patterns for rocksalt (NaCl) and wurtzite (ZnS), adjusted to lattice parameters of $a$=4.5 Å and $a$=3.3 Å/$c$=5.4 Å to approximate the average peak positions in the experimental patterns. The experimental XRD patterns match the peak positions and intensities of these high-symmetry structures well, with differences in relative intensities arising from both textured growth of the thin films, as well as different scattering powers within the unit-cell. Notably, we do not observe peak-splitting relative to the ideal wurtzite or rocksalt structures, suggesting disorder on the cation sites, which is not uncommon for nitrides deposited at low and moderate temperatures.[43] Further details of synthesis, structure prediction, and characterization can be found in **S.I.2**. Historically, the rate of discovering new ternary nitride $M_1$-$M_2$-N spaces has averaged ~3.3 per year, as illustrated in **Figure 2d**. Our rapid experimental realization of novel nitrides in 7 previously unexplored $M_1$-$M_2$-N spaces validates the predictions from the map, bolsters confidence in the 85 other predicted spaces with stable nitrides, and highlights the valuable role of computational materials discovery in accelerating exploratory synthesis in novel chemical spaces.

a)

| Wurtzite-Derived | Pred. Space Group | $\Delta H_f$ (eV/atom) |
|---|---|---|
| $Zn_2SbN_3$ | $Cmc2_1$ (#36) | -0.035 |
| $Zn_3WN_4$ | $Pmn2_1$ (#31) | -0.437 |
| $Zn_3MoN_4$ | $Pmn2_1$ (#31) | -0.367 |
| **Rocksalt-Derived** | **Pred. Space Group** | **$\Delta H_f$ (eV/atom)** |
| $Mg_2NbN_3$ | $C2/m$ (#12) | -1.17 |
| $MgHfN_2$ | $I4_1/amd$ (#141) | -1.66 |
| $MgZrN_2$ | $I4_1/amd$ (#141) | -1.52 |
| $MgTiN_2$ | $R$-3$m$ (#166) | -1.48 |

b)

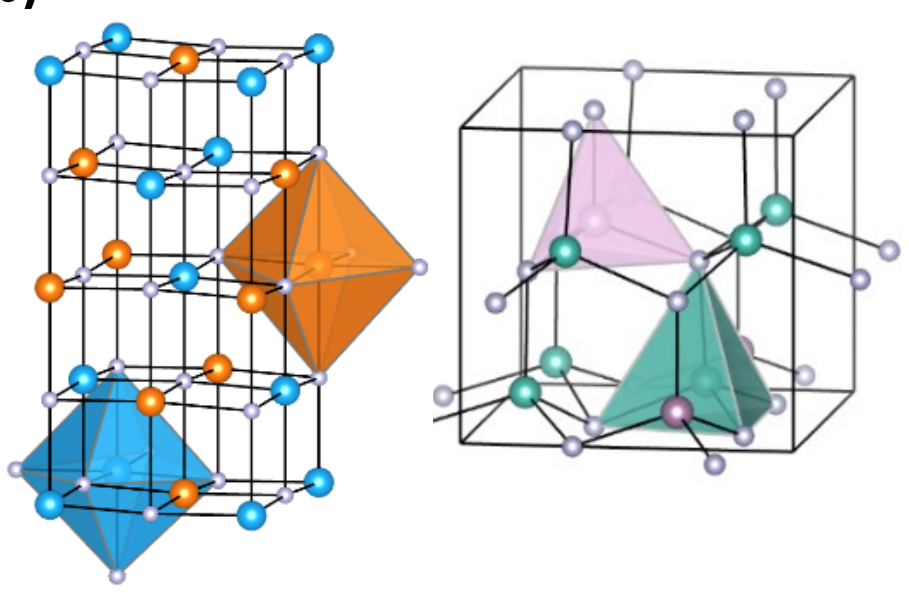

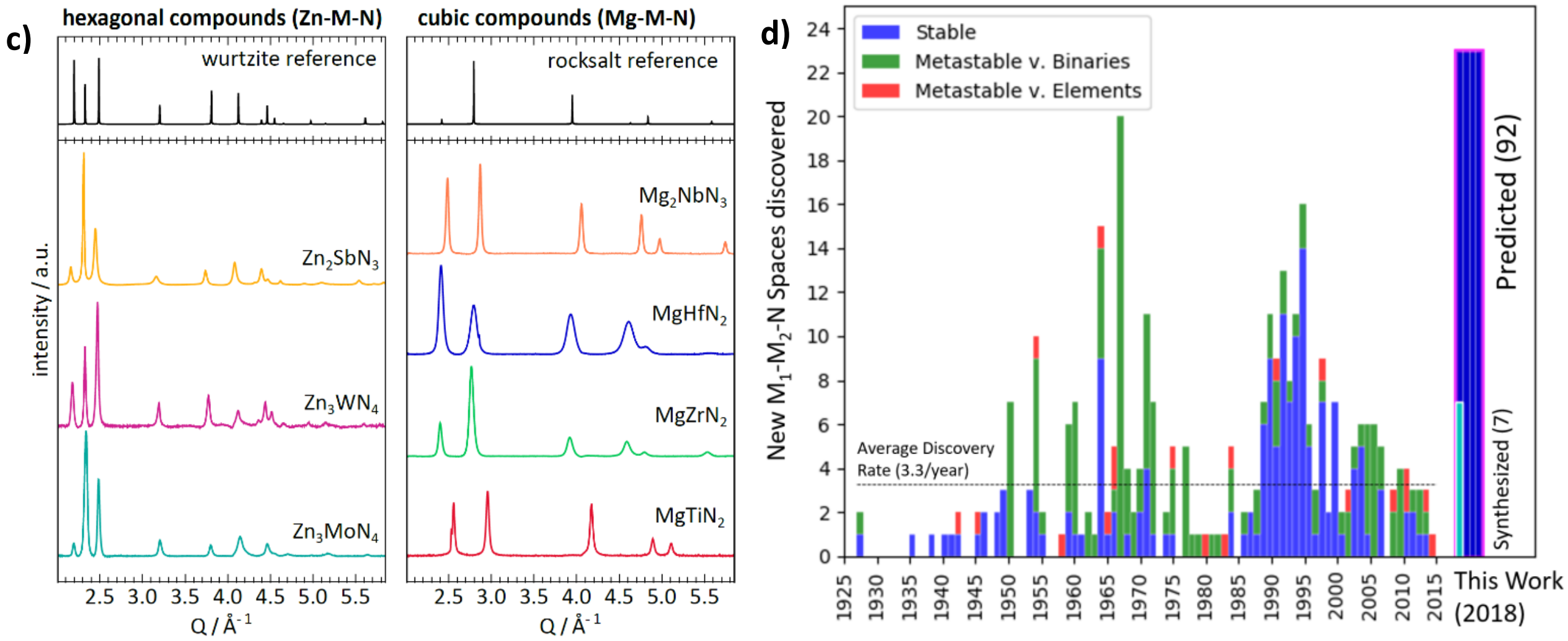

**Figure 2. a.)** Seven new Zn- and Mg- based ternary nitrides, with structures predicted by unconstrained structure search, and their corresponding formation energies. **b.)** Mg-based ternary nitrides are deposited in a rocksalt-derived structure (left), and Zn-based ternaries in a wurtzite-derived structure (right). **c.)** Synchrotron measured XRD patterns of new Zn- and Mg-based ternary nitrides, shown with reference diffraction patterns for lattice parameter-adjusted wurtzite and rocksalt. Q relates to diffraction angle ($\theta$) and incident wavelength ($\lambda$) by $Q = (4\pi/\lambda)\sin(\theta)$ and $\lambda = 0.9744$ Å **d.)** Discovery histogram for new ternary nitride spaces, based on entries as catalogued in the ICSD. Table of discovery dates in **S.I.4.**

**Metastable Ternary Nitrides**

Ternary nitrides that are metastable against decomposition into binary or elemental phases comprise the majority (71%) of the surveyed spaces. Most of the metastable nitrides with $\Delta H_f < 0$ (green) are mixed-transition metal nitrides, whereas mixed precious/post-transition metal nitrides typically have $\Delta H_f > 0$ (red). Although metastable nitrides should, in principle, be difficult to synthesize, ternary nitrides have been experimentally realized in 118 of the computed metastable spaces, shown in **Figure 1** by the inverted triangles. In our previous data-mining study of crystalline metastability,[44,45] we found nitrides to be the most metastable class of chemical compounds—having the largest fraction of metastable phases, as well as the highest average energies above the ground-state phases. The unusual metastability of nitrides can be attributed to the cohesivity afforded by strong metal-nitrogen bonds in the solid-state, which can kinetically 'lock-in' metastable nitride structures.

By formulating rational synthesis strategies to these metastable nitrides, we can expand the design space of functional nitride materials beyond equilibrium phases and compositions. One thermodynamic route to metastable nitrides is via nitrogen precursors that are less strongly bound than triple-bonded $N_2$;[36] such as ammonia,[17] azides,[46] or high-pressure supercritical $N_2$.[47] As an extreme example, plasma-cracked $N_2$ can yield atomic N precursors with nitrogen chemical potentials up to $\Delta\mu N \approx +1$ eV/N.[48,49] Thin-film synthesis from these precursors can form remarkably metastable nitrides, such as $SnTi_2N_4$ (metastable by 200 meV/atom),[50] or $ZnMoN_2$ in a wurtzite-derived structure (metastable by 160 meV/atom).[42] On **Figure 1**, we use orange boxes to highlight 95 spaces with metastable ternary nitrides predicted to be stabilizable under elevated nitrogen chemical potentials of $\Delta\mu N < +1$ eV/N.

Metastable ternary nitrides can also be obtained via soft solid-state synthesis routes; for example, delafossite $CuTaN_2$ is metastable by 127 meV/atom, but can be synthesized by ion-exchange metathesis of $Cu^+$ for $Na^+$ from the stable $NaTaN_2$ phase.[51] Amorphous precursors can also be a route to ternary nitrides that are metastable with respect to phase separation, whereby an atomically homogeneous amorphous precursor with the target ternary nitride composition is gently annealed to a lower energy, but still metastable, target crystalline phase.[52,53,54] On a separate note, decomposition of metastable ternary nitrides can also result in interesting functionality; for example, the segregation of metastable Si-Ti-N alloys at high temperature results in complex $TiN/Si_3N_4$ layered heterostructures with superior mechanical properties for tribological applications.[55]

# Discussion

By clustering the ternary nitrides space, we constructed a map that reveals broad overarching relationships between nitride chemistry and thermodynamic stability. We found alkali and alkaline earth ternary nitrides (Alk-Me-N) to comprise the majority of the stable ternary nitrides; whereas non-alkali metal-metal-nitrides (Me-Me-N) were generally found to be metastable against phase separation – albeit with some curious exceptions. Why do certain metals react favorably to form stable ternary nitrides, whereas others do not? Such questions probe at the very heart of solid-state chemistry.

We can achieve some insights towards this question by considering the geometric requirements of thermodynamic stability. A ternary nitride is stable if it is lower in free-energy than any stoichiometric combination of its competing ternaries, binaries, or elemental constituents. In formation energy versus composition space, this stability requirement manifests geometrically as a convex hull, illustrated for a ternary *A*-*B*-N space in **Figure 3a**. We can therefore rationalize the stability of a ternary nitride from 1) a thermochemical perspective–if a ternary nitride is lower in energy than its competing binary nitride(s), and 2) from a solid-state bonding perspective–how two metals interact electronically within a ternary nitride to raise or lower the bulk lattice energy of the ternary compound.

### Thermochemical decomposition into competing binaries

To quantify the thermochemical propensity of a ternary nitride to decompose into its competing binaries, we first define a feature named the 'depth of the binary hull', referring to the lowest energy binary nitride in a binary Me-N space. This binary hull depth, illustrated in **Figure 3a** by a black dashed line, serves as a proxy for the strength of the pairwise metal-nitrogen bond in the solid-state. A list of the 'deepest-hull' binaries and their formation energies are listed in **S.I.5**. We note that in some binary nitride spaces, the lowest formation-energy binary nitride has positive formation energy—for example, $Cu_3N$ in the Cu-N hull, indicating that $Cu_3N$ decomposes to Cu(s) + $N_2$(g) under ambient conditions.

**Figure 3b** shows for each element how many stable ternary spaces it forms in, versus the 'depth' of the binary hull. A volcano plot emerges, where elements that have either shallow or deep binary nitride hulls tend not to form many stable ternary nitrides, whereas elements that have intermediate binary nitride hull depths (around −0.8 eV/atom) form stable ternary nitrides most readily. From a thermochemical perspective, when the binary hull is deep, there is a greater propensity for a ternary metal nitride to phase-separate into its competing low-energy binary nitrides. On the other hand, a shallow (or positive) binary hull depth indicates intrinsically weak metal-nitrogen bonding; meaning ternary nitride formation is probably unfavorable in the first place. Intermediate binary hull depths indicate favorable metal-nitrogen bonding, but not enough for decomposition of a ternary nitride into its binary constituents, offering a compromise between these two competing effects.

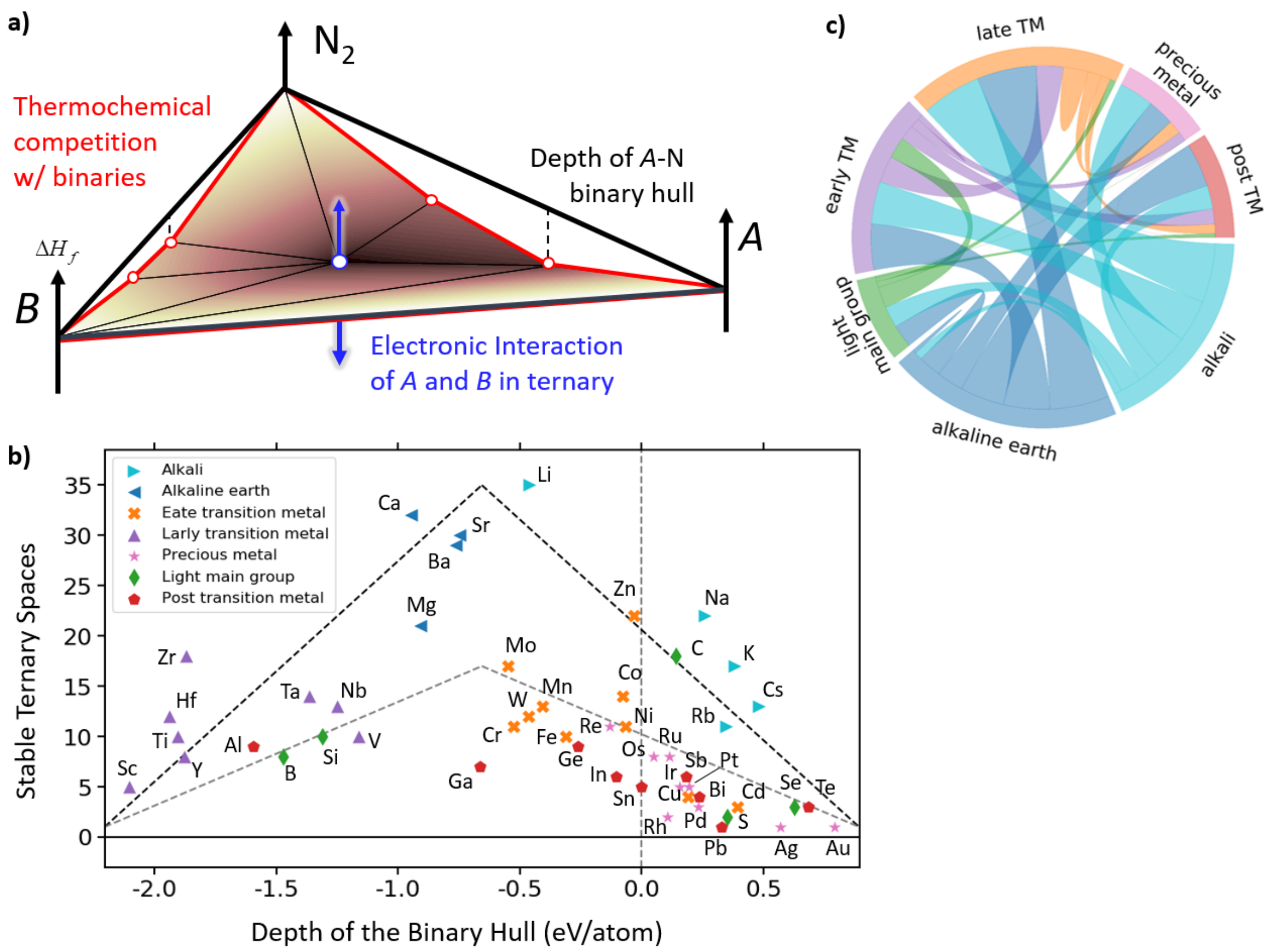

**Figure 3a.)** 3-D projection of a convex hull in a ternary *A*-*B*-N space, where the vertical axis is formation energy and the horizontal triangular plane is composition. The stability or metastability of a ternary nitride compound (blue circle) is governed by its propensity to decompose into competing binaries (red lines), as well as by its lattice energy arising from the electronic interaction of two metals in the ternary nitride (blue arrows). **b.)** Scatterplot showing the number of stable ternary systems each metal appears in, plotted against the depth of the binary hull, which corresponds to the formation energy of the lowest-energy binary nitride in the M-N binary space. Eye-guides for the volcano trend provided by dashed lines **c.)** Chord diagram showing the frequency of inter-group and intra-group bonding relationships of stable ternary nitrides.

In **Figure 3b**, the alkali and alkaline earth metals stand out on the volcano – forming stable ternaries more readily than other elements with similar binary hull depth. While the volcano plot captures the propensity of individual metals to decompose from a ternary nitride into their corresponding binaries, it does not capture how two metals influence each other electronically within the ternary nitride, which governs the bulk lattice energy of the ternary compound. **Figure 3c** visualizes in a chord diagram which periodic groups the two metals *A* and *B* in a stable ternary come from, where the width of each chord indicates the frequency of that combination. Our chord diagram shows that the two metals in a ternary compound tend to originate from different groups across the periodic table, where one of the elements is often an alkali or alkaline earth metal. This observation is consistent with heuristics arising from Hard-Soft / Acid-Base (HSAB) theory,[56] which suggests that ternary nitrides form most readily when the two acids (cations in ternary nitrides) have different HSAB character.[16]

## Electronic Origin of Ternary Nitride Stability

Qualitatively, we expect differences in electronegativity between *A*, *B* and N to redistribute the electron density into different bonds, which in the solid-state, may have mixed metallic, ionic and covalent character. Inspired by the role of 'conceptual DFT' in rationalizing the reactivity of atoms and molecules,[27,28] we construct new semi-quantitative schemes to extract the nature of solid-state bonding from the DFT-computed electron density. We compute ionic character of each ion from the ratio of the stoichiometrically-normalized Net Atomic Charges (NAC) over the Summed Bond Order (SBO) obtained from the Density Derived Electrostatic and Chemical (DDEC) approach.[57,58] We use Crystal Orbital Hamiltonian Population calculations[59] to decompose the integrated bonding energies of metal-metal interactions (*A*-*A*, *A*-*B*, *B*-*B*) as metallicity,[60] and non-metal interactions (*A*-N, *B*-N, N-N) as covalency. Using these features, we formulate data-driven insights into how this mixed solid-state bonding character influences ternary nitride stability.

We visualize our results on the classic metallic-ionic-covalent axes of van Arkel triangles shown in **Figure 4a**,[61] using hexagonal-binned histograms to represent the scatter distribution on these triangles. Full scatter plots for these van Arkel triangles can be found in **S.I.6**. From **Figure 4a**, we see that stable Alk-Me-N ternaries tend to exhibit greater ionicity and metal-nitrogen covalency, whereas stable Me-Me-N ternaries generally have higher metallicity. This distinction becomes even more apparent when the triangles are further separated by nitrogen-rich and nitrogen-poor nitrides, where this nitrogen excess or deficiency is compositionally referenced against the deepest-hull binary nitrides (details in **S.I.3**). For example, the formation of 'nitrogen-rich' $Ca_2VN_3$ from $Ca_3N_2$ and VN requires excess nitrogen, which is accompanied by the formal oxidation of vanadium from the binary nitride $V^{3+}N$ to the ternary $Ca_2V^{5+}N_3$. On the other hand, formation of 'nitrogen-poor' compositions occur by nitrogen release, such as the formation of $Co_2Mo_3N$ from CoN and MoN.

Stable Alk-Me-N are mostly nitrogen rich, whereas most stable Me-Me-N are nitrogen poor. This dichotomy between nitrogen-rich and nitrogen-poor ternary nitrides can largely be rationalized by how electron density redistributes between the nitrogen anion and the more electronegative metal during a reaction from the deepest-hull binaries to a stable ternary nitride. **Figure 4b** plots the changes in ionicity of the nitrogen anion, $\Delta\delta_N$, and the more electronegative metal cation, $\Delta\delta_B$, during such a reaction. The formation of nitrogen-rich nitrides typically involves *B*-metal oxidation and nitrogen reduction, whereas the formation of nitrogen-poor nitrides shows the opposite, exhibiting metal reduction and nitrogen oxidation. We emphasize that this nitrogen oxidation and reduction is measured relative to the nitrogen anion from the corresponding binary nitrides, not to the $N_2$ molecule.

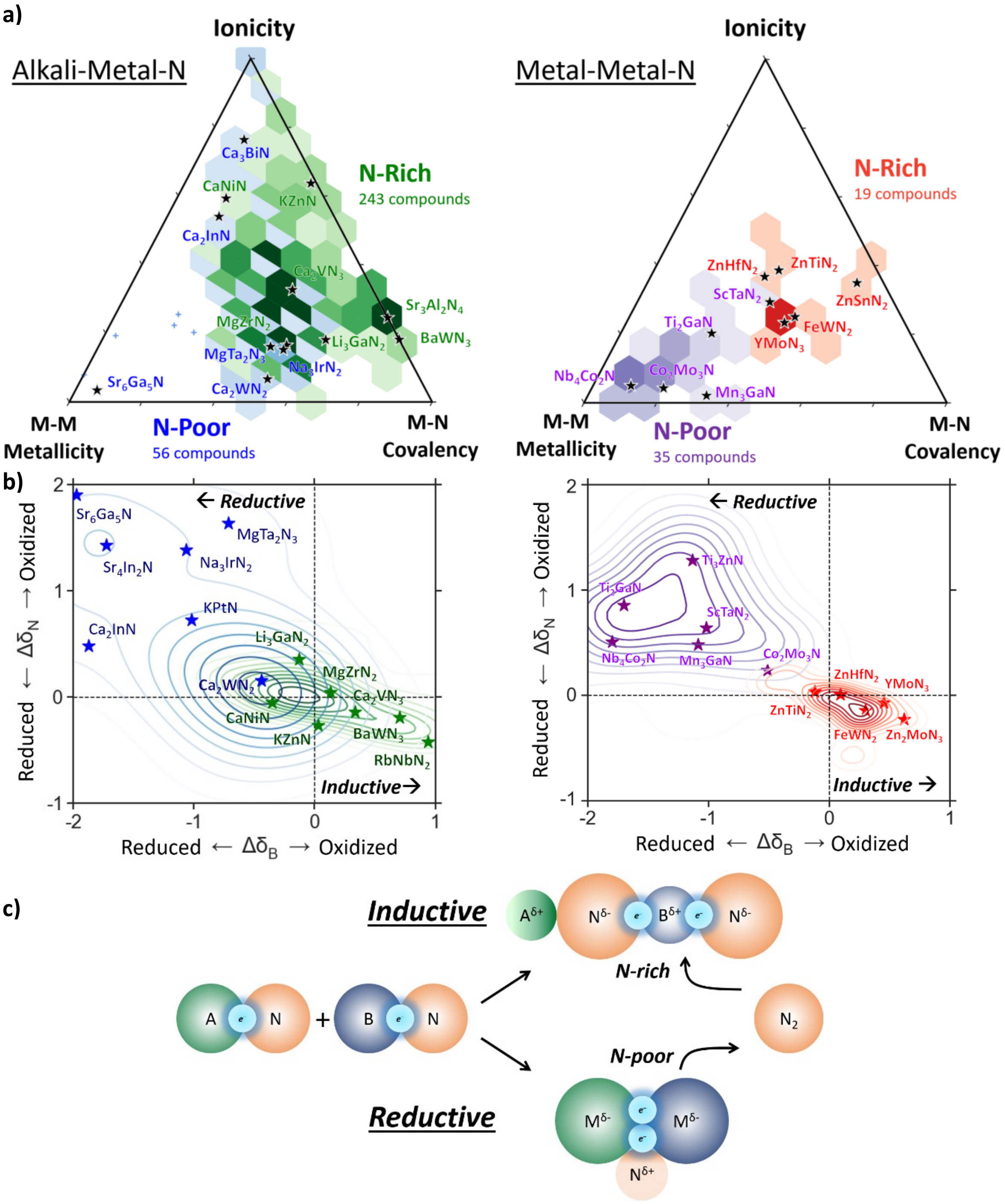

**Figure 4. a**.) Metallicity, ionicity, and covalency of the stable ternary nitrides; hexagonally binned on van Arkel triangles by the nitrogen-excess or nitrogen-deficiency of the ternary, compositionally-referenced against the deepest-hull binary nitrides. Hexagons plotted for regions with >2 data points only. Color intensity corresponds to number density in each hexagon. Outliers in the Alk-Me-N triangle are shown with small crosses. **b.)** Kernel density distributions of ion oxidation and reduction between a deepest-hull binary and the stable ternary nitride, for nitrogen (vertical axis) and the more electronegative metal, *B* (horizontal axis). **c.)** *Inductive effect*, electropositive metal A donates electron density to B-N covalent bond, oxidizing the more electronegative metal, which can lead to nitrogen-rich nitrides. *Reductive effect*, nitrogen oxidation or nitrogen release provides electrons to Me-Me bonds, reducing the metals and increasing metallicity.

The formation of nitrogen-rich nitrides can be rationalized primarily from the *Inductive Effect*,[1,62] where an electropositive metal, $A$, donates electron density to its adjacent nitrogen anion, driving the formation of strong nitrogen covalent bonds with the more electronegative metal, $B$. As illustrated in **Figure 4b** and **4c**, this electron donation from $A$ generally leads to nitrogen reduction, which in turn oxidizes the metal $B$. Significant oxidation of $B$ can be compensated by excess nitrogen—which explains the formation of nitrogen-rich nitrides. An oxidized cation and reduced anion increases the overall ionicity of the $A^{\delta+}[B\text{-N}]^{\delta-}$ framework, resulting in nitride ceramics with very negative electrostatic Madelung energies. Because alkali and alkaline metals are so electropositive, the inductive effect drives the strong exothermic formation energies of Alk-Me-N ternaries, rationalizing their predominance within the ternary nitrides map. The inductive effect can also be operative in nitrogen-rich Me-Me-N; most frequently with Zn, which is a relatively electropositive transition metal and can also serve as an electron donor. This fact was previously captured by the hierarchical agglomeration algorithm, which clustered Zn with the other alkali and alkaline earth metals.

For stable nitrogen-poor nitrides, we propose a novel *Reductive Effect*, where remarkably, nitrogen can serve as an electron donor for metal reduction. For some ternary Me-Me-N compositions, Me-Me bonds may be stronger than Me-N bonds. As shown in **Figure 4c**, the oxidation or release of electrophilic nitrogen atoms can redistribute electron density back to these strong Me-Me bonds, leading to the reduction of the corresponding metals. The reductive effect can stabilize unusual structures in the nitride chemistry;[63] for example, $Co_2Mo_3N$, which exhibits infinite 1-D chains of covalently-bonded $[\mathrm{Co}-\mathrm{Co}]_\infty$ intertwined within an extended Mo-N covalent network (structure in **Fig. S.I.7**). The reductive effect can also be operative in stable nitrogen-poor stoichiometries of Alk-Me-N compounds; for example, in $Sr_3Ge_2N_2$, which features infinite 1-D $[\mathrm{Ge}-\mathrm{Ge}]_\infty^{2-}$ chains throughout the otherwise ionic $(Sr^{2+})_2[GeN_2]^{4-}$ lattice (structure in **Fig. S.I.7**). The data-mining structure prediction algorithm used in this work operates on ionic substitution, which may not be ideally poised to predict novel nitrogen-poor nitrides due to their ambiguous valence states, suggesting there may still be many reductive effect-stabilized Me-Me-N ternary nitrides awaiting prediction.

Our analysis shows that the nitrogen anion can be fairly amphoteric in the solid-state—usually acting as an electron acceptor under the inductive effect to form ionic/covalent nitrogen-rich nitrides, but sometimes serving as an electron donor in the reductive effect, driving the formation of metallic nitrogen-poor sub-nitrides. The span of electronic structures available to the ternary nitrides offers a rich design space for materials functionality. Incorporating an alkali metal into an otherwise metallic binary nitride can increase charge localization driven by the inductive effect, opening a band gap and thus creating functional semiconducting nitrides suitable for solid-state lighting, photovoltaic energy conversion, piezoelectrics, and more. On the other hand, nitrogen-poor nitrides possess metallic bonding punctuated by charge-localization on nitrogen atoms, which can lead to complex electronic and magnetic structures[64] and may serve as the basis for novel superconductors and topologically-protected quantum materials.[65,66,67] Modifying the nitrogen stoichiometry within a chemical space can be an effective strategy to compositionally tune the electronic structure between the reductive and inductive effect. For example, varying the Zn/Mo ratio in a wurtzite-based Zn-Mo-N compound can modulate the molybdenum oxidation state from $Mo^{4+}$ to $Mo^{6+}$, turning conductive $ZnMoN_2$ into insulating $Zn_3MoN_4$, a wide-bandgap semiconductor.[42]

## Conclusions

The library of inorganic solids has been dominated by oxides, whose structures and chemistries are often known from mineralogy. Compounds that do not form readily under ambient conditions, such as nitrides, offer a new frontier for materials discovery and design—so long as we have a rational understanding of the factors that drive stability in these relatively unexplored spaces. In this work, we used computational materials discovery and informatics tools to build a large stability map of the ternary metal nitride space. Our objective was not only to predict and synthesize new ternary metal nitrides, but further, to visualize large-scale relationships between nitride chemistry and thermodynamic stability, and to rationalize these trends from their deeper chemical origins. Our map as it stands is necessarily incomplete—it represents a current 'upper-bound' on the ternary nitride stability landscape. As new exotic structures and bonding motifs are discovered in the ternary metal nitrides, the procedures in this work can be iteratively re-applied to update and refine our understanding of this extended compositional space. From a broader perspective, our computational approach offers a systematic blueprint for mapping uncharted chemical spaces, providing synthetic chemists guidance in their quest to continuously extend the frontier of solid-state chemistry.

## ***Acknowledgements***

Funding for this study was provided by the US Department of Energy, Office of Science, Basic Energy Sciences, under Contract no. UGA-0-41029-16/ER392000 as a part of the DOE Energy Frontier Research Center "Center for Next Generation of Materials Design: Incorporating Metastability." This research used resources of the Center for Functional Nanomaterials, which is a U.S. DOE Office of Science Facility, at Brookhaven National Laboratory under Contract No. DE-SC0012704. This work also used computational resources sponsored by the Department of Energy's Office of Energy Efficiency and Renewable Energy, located at NREL. C.J.B and A.M.H. acknowledge support in part from the Research Corporation for Science Advancement through the Scialog: Advanced Energy Storage award program. Use of the Stanford Synchrotron Radiation Lightsource, SLAC National Accelerator Laboratory, is supported by the U.S. Department of Energy, Office of Science, Office of Basic Energy Sciences under Contract No. DE-AC02-76SF00515.

[26] Butler, Keith T., et al. "Machine learning for molecular and materials science." *Nature* 559.7715 (2018): 547.

[27] Parr, Robert G., and Weitao Yang. "Density functional approach to the frontier-electron theory of chemical reactivity." *Journal of the American Chemical Society* 106.14 (1984): 4049-4050.

[28] Hoffmann, Roald. "How chemistry and physics meet in the solid state." *Angewandte Chemie International Edition in English* 26.9 (1987): 846-878.

[29] Hinuma, Yoyo, et al. "Discovery of earth-abundant nitride semiconductors by computational screening and high-pressure synthesis." *Nature communications* 7 (2016): 11962.

[30] Tholander, Christopher, et al. "Strong piezoelectric response in stable TiZnN2, ZrZnN2, and HfZnN2 found by ab initio high-throughput approach." *Journal of Applied Physics* 120.22 (2016): 225102.

[31] Gharavi, Mohammad Amin, et al. "Theoretical study of phase stability, crystal and electronic structure of MeMgN2 (Me= Ti, Zr, Hf) compounds." *Journal of Materials Science* 53.6 (2018): 4294-4305.

[32] Ching, W. Y., et al. "Prediction of spinel structure and properties of single and double nitrides." *Physical Review B*63.6 (2001): 064102.

[33] Sarmiento-Perez, Rafael, et al. "Prediction of stable nitride perovskites." *Chemistry of Materials* 27.17 (2015): 5957-5963.

[34] Bannikov, V. V., I. R. Shein, and A. L. Ivanovskii. "Elastic properties of antiperovskite-type Ni-rich nitrides MNNi3 (M= Zn, Cd, Mg, Al, Ga, In, Sn, Sb, Pd, Cu, Ag and Pt) as predicted from first-principles calculations." *Physica B: Condensed Matter* 405.22 (2010): 4615-4619.

[35] Hautier, Geoffroy, et al. "Data mined ionic substitutions for the discovery of new compounds." *Inorganic chemistry* 50.2 (2010): 656-663.

[36] Sun, Wenhao, et al. "Thermodynamic Routes to Novel Metastable Nitrogen-Rich Nitrides." *Chemistry of Materials*29.16 (2017): 6936-6946.

[37] Ong, Shyue Ping, et al. "Python Materials Genomics (pymatgen): A robust, open-source python library for materials analysis." *Computational Materials Science* 68 (2013): 314-319.

[38] Jain, A., Ong, S.P., Hautier, G., Chen, W., Richards, W.D., Dacek, S., Cholia, S., Gunter, D., Skinner, D., Ceder, G. and Persson, K.A., Commentary: The Materials Project: A materials genome approach to accelerating materials innovation. *APL Mater.* **2013,** 1, 011002.

[39] Kemp, Charles, and Joshua B. Tenenbaum. "The discovery of structural form." *Proceedings of the National Academy of Sciences* 105.31 (2008): 10687-10692.

[40] Müllner, Daniel. "Modern hierarchical, agglomerative clustering algorithms." *arXiv preprint arXiv:1109.2378* (2011).

[41] Gower, John C. "A general coefficient of similarity and some of its properties." *Biometrics* (1971): 857-871.

[42] Arca, Elisabetta, et al. "Redox-Mediated Stabilization in Zinc Molybdenum Nitrides." *Journal of the American Chemical Society* 140.12 (2018): 4293-4301.

[43] Lany, Stephan, et al. "Monte Carlo simulations of disorder in ZnSn N 2 and the effects on the electronic structure." *Physical Review Materials* 1.3 (2017): 035401.

[44] Sun, Wenhao, et al. "The thermodynamic scale of inorganic crystalline metastability." *Science Advances* 2.11 (2016): e1600225.

[45] Bartel, Christopher J., et al. "Physical descriptor for the Gibbs energy of inorganic crystalline solids and prediction of temperature-dependent materials chemistry." *arXiv preprint arXiv:1805.08155* (2018).

[46] Choi, Jonglak, and Edward G. Gillan. "Solvothermal metal azide decomposition routes to nanocrystalline metastable nickel, iron, and manganese nitrides." *Inorganic chemistry*48.10 (2009): 4470-4477.

[47] Kroll, P., Schröter, T., Peters, M. Prediction of novel phases of tantalum (V) nitride and tungsten (VI) nitride that can be synthesized under high pressure and high temperature. *Angew. Chem., Int. Ed.* **2005,** 44.27, 4249-4254.

[48] Caskey, Christopher M., et al. "Thin film synthesis and properties of copper nitride, a metastable semiconductor." *Materials Horizons* 1.4 (2014): 424-430.

[49] Caskey, Christopher M., et al. "Synthesis of a mixed-valent tin nitride and considerations of its possible crystal structures." *The Journal of chemical physics* 144.14 (2016): 144201.

[50] Bikowski, Andre, et al. "Design of Metastable Tin Titanium Nitride Semiconductor Alloys." *Chemistry of Materials* 29.15 (2017): 6511-6517.

# A Map of the Inorganic Ternary Metal Nitrides

Wenhao Sun[1], Christopher Bartel[2], Elisabetta Arca[3], Sage Bauers[3], Bethany Matthews[4], Bernardo Orvañanos[5], Bor-Rong Chen,[6] Michael F. Toney,[6] Laura T. Schelhas,[6] William Tumas[3], Janet Tate,[4] Andriy Zakutayev[3], Stephan Lany[3], Aaron Holder[2,3], Gerbrand Ceder[1,7]

[1] Materials Sciences Division, Lawrence Berkeley National Laboratory, Berkeley, California 94720, USA

[2] Department of Chemical and Biological Engineering, University of Colorado, Boulder, Colorado 80309, USA

[3] National Renewable Energy Laboratory, Golden, Colorado 80401, USA

[4] Department of Physics, Oregon State University, Corvallis, Oregon 97331, USA

[5] Department of Materials Science and Engineering, Massachusetts Institute of Technology, Cambridge, MA 02139

[6] SLAC National Accelerator Laboratory, Menlo Park, CA, 94025, USA.

[7] Department of Materials Science and Engineering, UC Berkeley, Berkeley, California 94720, USA

## Supplemental Information

# S.I.1 Interactive map of ternary metal nitrides

To view the interactive ternary nitrides map, extract the attached **InteractiveTernaryNitridesMap.zip** file into a folder, and open **TernaryNitridesMap.html**. Hovering the mouse cursor over an individual $M_1$-$M_2$-N entry on the ternary nitrides map will show the corresponding ternary nitride phase diagram, along with a table of stable and metastable entries. An example is shown below for the Zn-Ti-N system.

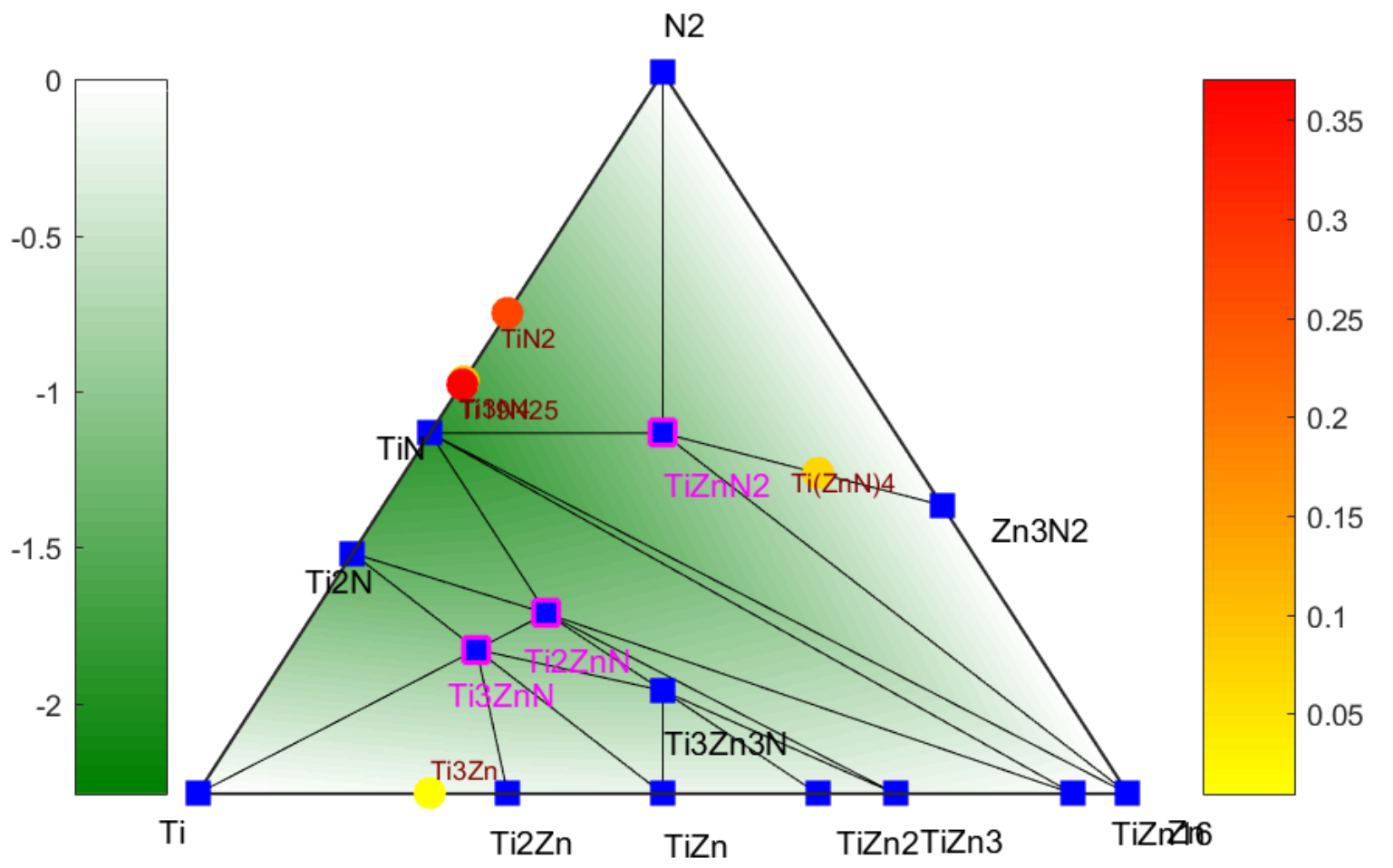

| Stable | Known? | Formation Enthalpy (eV/atom) | Energy below hull (eV/atom) |
|---|---|---|---|
| Ti3ZnN | New | -0.972 | -0.031 |
| Ti3Zn3N | ICSD | -0.755 | -0.012 |
| TiZnN2 | New | -1.036 | -0.069 |
| Ti2ZnN | New | -1.154 | -0.035 |

| Metastable | Known? | Formation Enthalpy | Energy abv hull | Decomposition | Critical $\mu_N$ (eV/N) |
|---|---|---|---|---|---|
| Ti(ZnN)4 | New | -0.409 | 0.07 | TiZnN2+Zn3N2 | 0.07 |

The green background of the ternary phase diagram corresponds to the depth of the ternary convex hull as a function of composition; color coded by the formation energy as shown on the left colorbar (units of eV/atom). Stable phases are indicated on the ternary phase diagram by a blue square. New stable ternaries predicted in this work are indicated by a magenta box (in this example, $Ti_3ZnN$, $Ti_2ZnN$, $TiZnN_2$). Stable ternaries without a magenta box (such as $Ti_3Zn_3N$) can be found in the ICSD. The 'energy below hull' indicates the reaction energy of a stable ternary nitride from its stable neighbors in phase space.

Metastable compounds are indicated by a circle, with the energy above the hull indicated by the color bar on the right, in units of eV/atom. To avoid clutter, we only include metastable compounds with an energy above the hull < 400 meV/atom. Note that we previously found the $90^{th}$ percentile of nitride metastability to be 200 meV/atom.[1] If there are no ternary nitrides within 400 meV/atom of the hull, then the lowest formation energy metastable ternary is listed on the table below. The table also includes the decomposition products of a metastable ternary nitride. For metastable ternary nitrides that decompose to $N_2$, we also provide a critical $\mu_N$ that can stabilize these metastable compounds, see Ref 2 for thermodynamic details.

The most up-to-date crystal structures for these compounds, along with their computed properties, can be obtained by searching for these compositions on the Materials Project.

The interactive map was programmed in Bokeh, an interactive python visualization library.[3] Buttons on the bottom of the map can be used to turn on/off the ICSD triangles, the hover feature for phase diagram information, and the crosshair feature. Zooming in or out of the map using your internet browser (Ctrl + '=/–') may facilitate your viewing of the map. We recommend using the Google Chrome browser.

# S.I.2 Methodology

## S.I.2.a Computational Methods

### 1. Data-Mined Ionic Substitution Structure Prediction

Ternary structures were generated using the Data-Mined Structure-Prediction algorithm (DMSP), with methodology described in References 4,5, and trained specifically for nitride discovery in Reference 6. Briefly, the substitution training matrix was trained on the Inorganic Crystal Structure Database (ICSD), mapping isostructural compounds and identifying which cations are statistically probable to substitute for one another. The training set for nitride discovery was performed on the set of all pnictides (N + P + As + Sb), which has superior prediction capability to an oxide-containing set. This training set was found via 10-fold cross validation to have 80% probability of recovering known nitride structures. Training of the DMSP algorithm was performed on the ICSD as extracted in October 2015.

### 2. Density Functional Theory Calculations

Total energies of known and DMSP-suggested nitrides were calculated with density functional theory using the Vienna *ab initio* software package (VASP),[7,8] using the projector augmented-wave method with the PBE exchange-correlation functional and PAW pseudopotentials in VASP. Plane-wave basis cut-off energies are set to 520 eV. The k-point densities were distributed within the Brillouin zone in a Monkhorst-Pack grid,[9] or on a Gamma-centered grid for hexagonal cells, and used default *k*-point densities in compliance with Materials Project calculation standards,[10] which were calibrated to achieve total energy convergence of better than 0.5 meV/atom. Each structure is initiated in ferromagnetic, ferrimagnetic, and anti-ferromagnetic spin configurations, and the lowest-energy configuration is used for phase stability calculations.

Phase stability calculations are computed from convex hulls, using the phase diagram analysis package in Pymatgen,[11] calculated with respect to known nitride phases from the Materials Project,[12] obtained using the Materials Project REST API.[13] Azides (e.g. $NaN_3$, $WN_{18}$) are removed from the phase diagram when computing phase stability, as they do not typically form during solid-state synthesis techniques. Materials Project data was retrieved in January, 2018.

For the Zn-Mo-N, Zn-W-N, Zn-Sb-N, Mg-Ti-N, Mg-Zr-N, Mg-Hf-N, Mg-Nb-N systems, an unconstrained ground-state search was performed using the "Kinetically Limited Minimization" approach,[14] which does not require prototypical structures from databases. Seed structures are generated from random lattice vectors and atomic positions, subject to geometric constraints to avoid extreme cell shapes, and to observe minimal interatomic distances (2.8 Å for cation-cation and anion-anion pairs, 1.9 Å for cation-anion pairs). For each material, we sampled at least 100 seeds, over the ternary compositions $A_iB_jN_k$ for $ijk$ = 112, 146, 414, 213, 124, 326, 338, 313, chosen to accommodate the $(Mg/Zn)^{2+}$, $M^{4+/5+/6+}$, and $N^{3-}$ oxidation states. New trial structures are generated by the random displacement of one atom between 1.0 and 5.0 Å while maintaining the minimal distances. Trial structures are accepted if the total energy is lowered, and the number of trials equals the number of atoms in the unit cell.

### 3. Extracting chemical insights from first-principles calculations

The Density Derived Electrostatic and Chemical (DDEC) approach was used to obtain net atomic charges and natural bond orders assigned to each ion in each calculated structure.[15,16] From the DDEC analysis we define the average charge for ion $i$, $\delta_i$, as the net atomic charge assigned to ion, $i$ (number of electrons) averaged over all ions, $i$, in the structure, $A_\alpha B_\beta N_\gamma$. The summed bond order for ion $i$, $s_i$, was obtained similarly by summing the natural bond orders for all interactions containing $i$, averaged over all ions, $i$, in the structure, $A_\alpha B_\beta N_\gamma$. The Crystal Orbital Hamilton Population (COHP) analysis was used to quantify the bonding interactions within each structure and partition these interactions by specific ion-ion pairs using the LOBSTER code.[17] To normalize the comparison of COHPs across a range of structures and compositions, the energy levels from each PBE calculation were aligned to core levels. Doing so allows for a reasonable comparison of Fermi energies and thus COHP energy depths across the various systems analyzed. To alleviate the effects of varied pseudopotentials across systems, the number of free atom valence electrons was determined for each system using the following equation:

$$N_v\left(A_\alpha B_\beta N_\gamma\right) = \alpha N_v(A) + \beta N_v(B) + \gamma N_v(N)$$

where $N_v$ is the number of valence (outermost shell) electrons. The minimum energy which contains valence electrons, $\varepsilon_V$ was then determined for each structure by incrementally decreasing the energy, $\varepsilon$, and integrating the density of states (DOS) from $\varepsilon$ to the Fermi energy, $\varepsilon_F$, such that

$$\varepsilon_V = \varepsilon_V\colon \int_{\varepsilon_V}^{\varepsilon_F} DOS(E')\, dE' = 1$$

where DOS is normalized by $N_v$ and $E'$ is the core-level aligned energy. The magnitude of bonding interactions, $\Sigma$, in each structure is then defined as

$$\Sigma = \int_{\varepsilon_V}^{\varepsilon_F} -COHP(E')E'\, dE'$$

where the COHP is also normalized by $N_v$.

Using these quantities, we produced the triangle plots shown in **Fig. 4a**. The ionicity, $I$, was defined as:

$$I_{A_\alpha B_\beta N_\gamma} = \frac{1}{\alpha+\beta+\gamma}\left(\alpha\frac{\delta_A}{s_A} + \beta\frac{\delta_B}{s_B} + \gamma\frac{\delta_N}{s_N}\right)$$

and quantifies the extent of electron transfer in the structure. The metallicity was defined as:

$$M_{A_\alpha B_\beta N_\gamma} = |\Sigma_{A-B} + \Sigma_{A-A} + \Sigma_{B-B}|$$

quantifying the net bonding energy of cation-cation interactions. The covalency, $C$, was defined as:

$$C_{A_\alpha B_\beta N_\gamma} = |\Sigma_{A-N} + \Sigma_{B-N} + \Sigma_{N-N}|$$

quantifying the net bonding energy for interactions containing nitrogen. To ensure each quantity ($C$, $I$, $M$) was of the same magnitude, each quantity was divided by the maximum of that quantity within the dataset. In order to plot points on a triangle, the sum of each point, (*C*, *I*, *M*), must equal 1. Therefore, each quantity within each point was normalized by $C+I+M$.

To quantify the extent to which a given ternary was "N-rich" or "N-poor", we compare the cation/anion ratios in the ternary to the ratios in the deep-hull binaries ($A_\alpha$-N, $B_\beta$-N) using the assumed formation reaction:

$$\left(\frac{\alpha}{\alpha'}\right) A_{\alpha'}N + \left(\frac{\beta}{\beta'}\right) B_{\beta'}N + 0.5\left(\gamma - \frac{\alpha}{\alpha'} - \frac{\beta}{\beta'}\right) N_2 \rightarrow A_\alpha B_\beta N_\gamma$$

and subsequent condition for being rich or poor in nitrogen:

$$\gamma - \frac{\alpha}{\alpha'} - \frac{\beta}{\beta'} \geq 0 \rightarrow Nrich;\ \gamma - \frac{\alpha}{\alpha'} - \frac{\beta}{\beta'} < 0 \rightarrow Npoor.$$

This reaction was also used to compute the change in charge, Δδ, across this formation reaction, as shown in **Fig. 4b**:

$$\Delta\,\delta_i = \delta_{i,A_\alpha B_\beta N_\gamma} - \frac{2}{\frac{\alpha}{\alpha'} + \frac{\beta}{\beta'} + \gamma}\left(\frac{\alpha}{\alpha'}\delta_{i,A_{\alpha'}N} + \frac{\beta}{\beta'}\delta_{i,B_{\beta'}N} + 0.5\left(\gamma - \frac{\alpha}{\alpha'} - \frac{\beta}{\beta'}\right)\delta_{i,N_2}\right)$$

where $\delta_{i,N_2}$ was taken to be 0. *A* and *B* are defined as the least and most electronegative cations in the ternary.

## S.I.2.B. Experimental Methods

### 1. Experimental Synthesis

Experimental synthesis of Zn-Me-N and Mg-Me-N thin-film sample libraries was carried out by combinatorial RF magnetron sputtering using high purity metals targets as cations sources. Ar and $N_2$ gas or Ar and activated $N_2$ by means of a nitrogen plasma source were used as sputtering gases. Two sputtering guns were loaded with 2"-diameter metal targets and positioned at 45° with respect to the substrate normal to achieve a cation compositional gradient; the N source was aligned to the normal. Prior to deposition, the sputtering chambers were evacuated to a pressure lower than $3\times10^{-6}$ Torr. Eagle-XG glass or fused silica slides were used as substrates. Before the deposition, the substrates were cleaned by degreasing in water and soap, followed by ultrasonic cleaning in acetone and isopropanol baths. Depositions were performed at a variety of pressure, gas flows, temperature, sputtering powers on the guns to achieve the desired crystalline phases. The deposition conditions used for the samples reported in this contribution are summarized on Table 1.

| | **Mg-Ti-N** | **Mg-Zr-N** | **Mg-Hf-N** | **Mg-Nb-N** |
|---|---|---|---|---|
| **Dep. Temp** | 500 °C | | | |
| **Substrate** | EXG | EXG | EXG | f-SiO2 |
| **$N_2$ flow** | 6 sccm | 6 sccm | 6 sccm | 6 sccm |
| **Ar flow** | 6 sccm | 6 sccm | 6 sccm | 6 sccm |
| **Deposition pressure** | 3.5 mT | 3.5 mT | 3.5 mT | 3.5mT |
| **Mg power** | 40 W | 40 W | 30 W | 40 W |
| **TM power** | 60 W | 60 W | 50 W | 40 W |
| **$N_2$ power** | 350 W | 350 W | 300 W | 350 W |
| **Dep time** | 90 mins (150-300nm thick depending on composition) | | | |

| | **Zn-Mo-N** | **Zn-W-N** | **Zn-Sb-N** |
|---|---|---|---|
| **Dep. Temp** | 200 | RT | 300 |
| **Substrate** | EXG | EXG | EXG |
| **$N_2$ flow** | 10 sccm | 10 sccm | 10 sccm |
| **Ar flow** | 10 sccm | 10 sccm | 10 sccm |
| **Deposition pressure** | 20 mT | 20 mT | 20 mT |
| **Zn power** | 30 | 35 W | 50 |
| **M power** | 40 W | 35 W | 20 W |
| **$N_2$ power** | 250 W | - | - |
| **Dep time** | 60 mins (100-500nm thick depending on composition) | | |

## 2. Experimental characterization

Spatially resolved characterization of the thin-film sample libraries was performed on a 4×11 grid consisting of 4 rows and 11 columns, with the compositional gradient along the 11 positions in one row. Cation composition was determined by quantitative XRF on Fischer Scientific instruments (Fischerscope XDV-SDD or Fischerscope XUV-XRF). Stylus profilometry (Dektak 8) or calibrated XRF were used to determine sample thickness. Data analysis was performed using custom-written procedures in the Wavemetrics IGOR Pro software package. The resulting data are available at htem.nrel.gov. Based on the results of the compositional analysis, sample points having the exact cation stoichiometry as to the theoretically predicted compounds, were subjected to high resolution x-ray diffraction measurements.

Wide-angle X-ray scattering (WAXS) was performed using the Stanford Synchrotron Radiation Lightsource (SSRL) Beamline 11–3 with an X-ray wavelength of 0.9744 Å. Two-dimensional scattering data were collected using a Rayonex MX225 detector in a grazing incidence geometry with the X-ray beam held at an incident angle of 3° and a sample to detector distance of 150 mm. Images are calibrated using a LaB6 standard and integrated between 10º < χ < 170º (χ is the polar angle) using GSAS-II.[18] Diffraction data presented in the main text were background subtracted using an ad-hoc Chebyschev function. Raw integrated line scans are shown in **Fig. S.I.2.1**. The broad signals at ca. 1.8 and 5 $Å^{-1}$ is from the amorphous substrates.

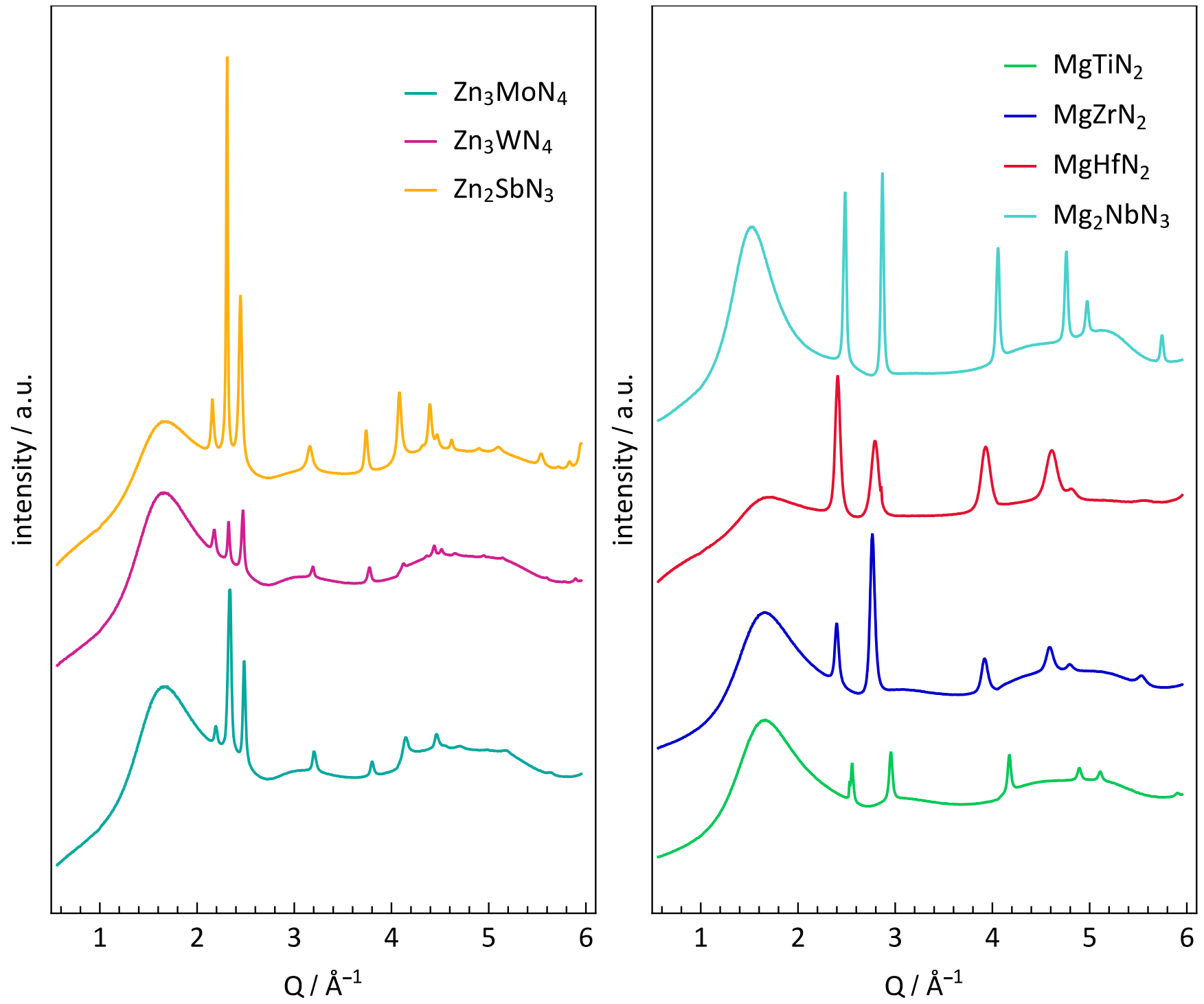

**Figure S.I.2.1** Unprocessed line-scans generated from integrated detector images for (left) hexagonal Zn-containing compounds and (right) cubic Mg-containing compounds. Q relates to diffraction angle ($\theta$) and incident wavelength ($\lambda$) by $Q = (4\pi/\lambda)\sin(\theta)$ and $\lambda = 0.9744$ Å

## S.I.3 Construction and Clustering of Ternary Nitrides Map

The stability information of a ternary chemical space can be represented on a heat map, constructed using axes corresponding to cations, and with a color scale corresponding to the formation energy of the lowest-energy compound in a ternary space. For a heat map with *N* elements on the axis, there are *N*! (N factorial) possible cation orderings of the heat map axis. Simple ordering schemes can be chosen corresponding to atomic number, Mendeleev number, or another predefined elemental ordering derived from the Periodic Table. However, because these axes may not be ordered specifically with respect to underlying relationships within in a chemical space, stability trends and chemical families may not be readily apparent. For example, shown below is a heat map for the stable ternary nitrides, with elements on the axes ordered by atomic number:

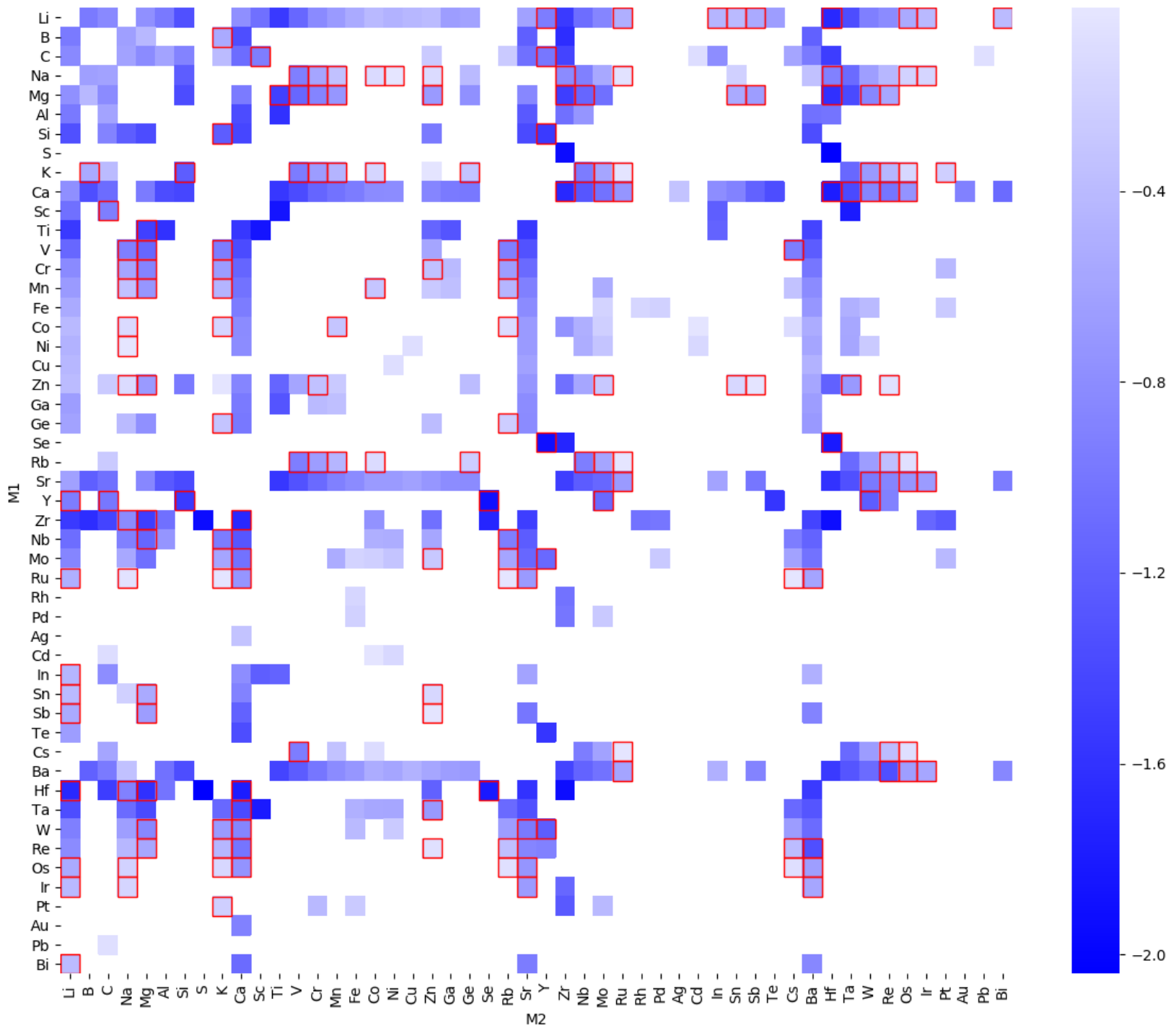

**Figure S.I.3.1** Unclustered heatmap, where blue color scale corresponds to the formation energy of the 'deepest-hull' stable ternary nitride within a ternary $M_1$-$M_2$-N space.

Although some relationships between elements are apparent by identification, for example between Li, Ca, Sr, and Ba, other relationships are difficult to discern by eye. To construct a visualization that places 'similar' elements adjacent to one another, we use hierarchical agglomeration algorithms to cluster the ternary nitride space. Clustering requires a distance metric, represented by a vector for each element, *A*, indicating its 'distance' to all other elements, *B*. In available python packages, such as *scipy* or Seaborn, this distance is usually given by a single feature, for example, formation energy. Using this distance feature, a clustering algorithm iteratively determines which elements are 'nearby' other elements, gradually building larger and larger clusters. These clusters, and the distances between them, can be represented phenotypically by a dendrogram. Single-feature cluster maps are straightforward to construct, for example, using the *clustermap* function in the Seaborn package. Shown below are several single-feature clustermaps:

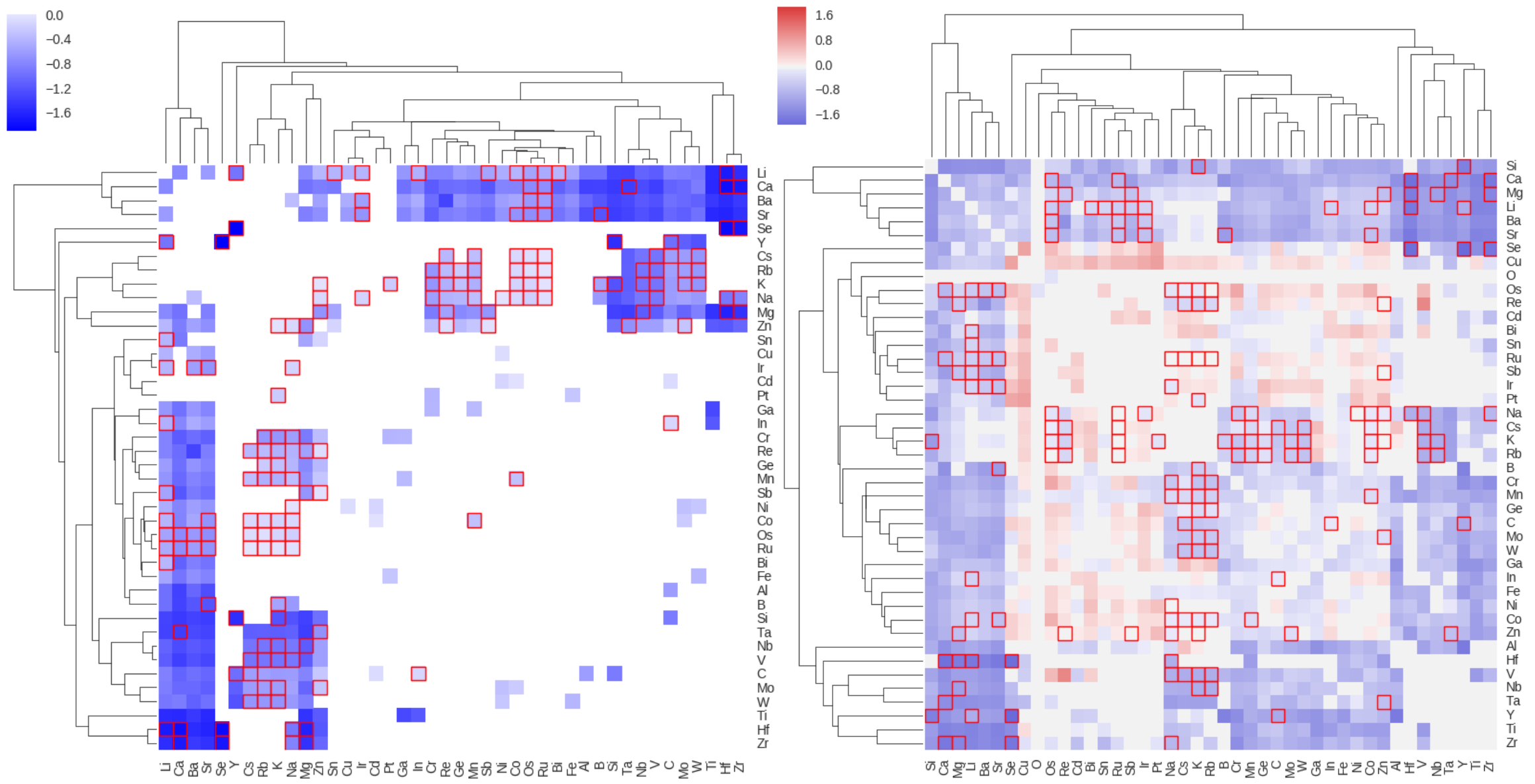

**Figure S.I.3.2.** Heatmaps clustered on the lowest formation-energy ternary nitride in a chemical space (*continuous data*) for **Left.)** stable ternary nitrides only, and **Right.)** including metastable ternaries.

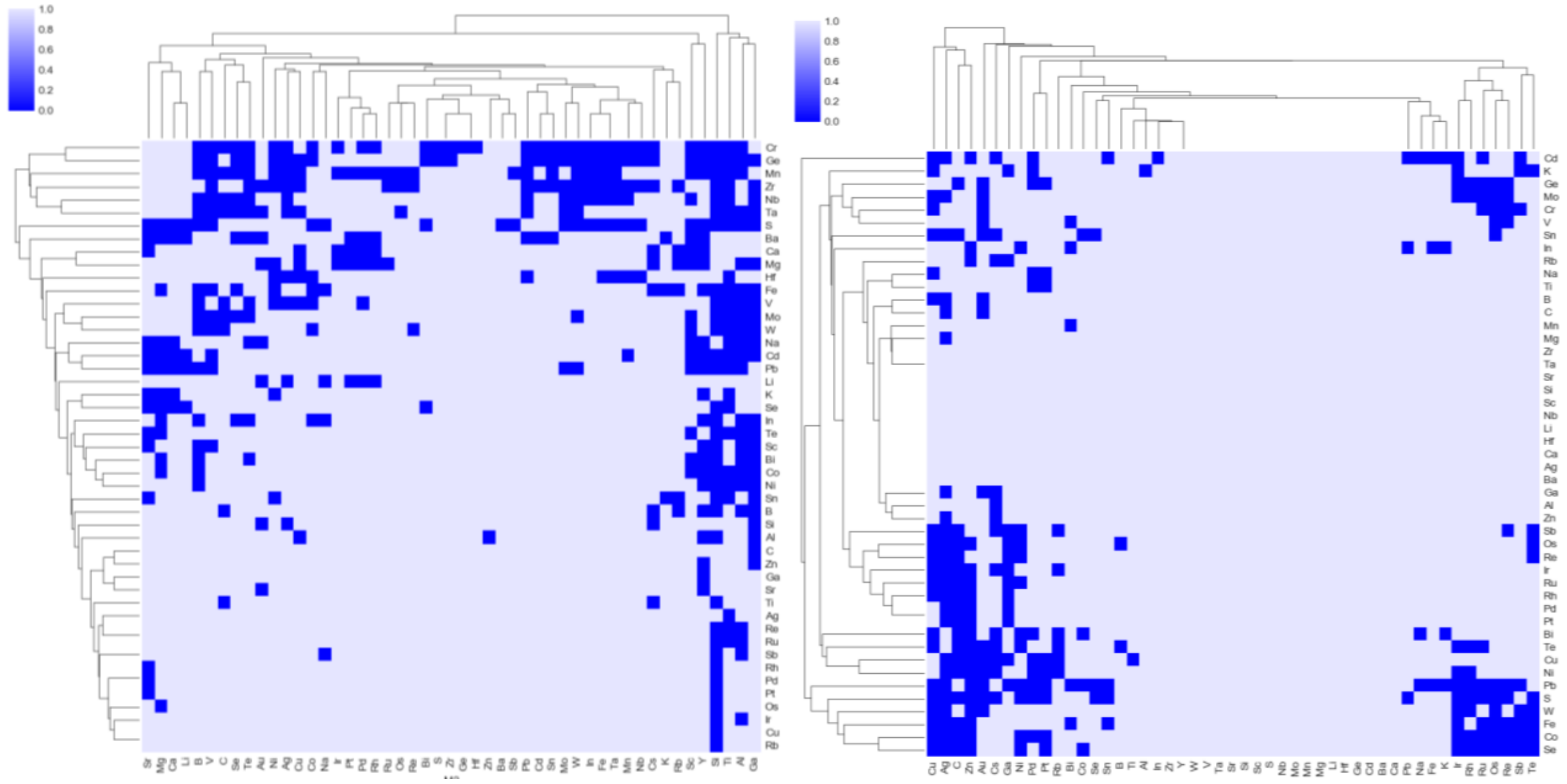

**Figure S.I.2.3.** Heatmaps clustered on nominal data, on whether the deepest compound in a ternary nitride space is **Left.)** metastable with respect to binaries (eventually on map as green), or **Right.)** metastable with respect to the elements (eventually on map as red).

For each singular feature, a dendrogram ordering and corresponding clustermap can be constructed. However, single-feature clustermaps present several limitations: 1) relationships constructed from different distance-metrics are difficult to cross-reference between maps, 2) dendrogram distances between elements with sparse features are not particularly meaningful, and 3) multiple relationships cannot be represented on the same figure.

To circumvent these limitations, we constructed a new multi-feature distance metric designed to capture multiple aspects of thermodynamic stability within the same figure. For the ternary nitrides map, our overarching objective is to elucidate which families of elements have similar thermodynamic propensity for forming ternary nitrides. At a high level, we aim to distinguish the elements by their propensity to form stable or metastable nitrides, and at a more local level, to group these elements by their formation energies and periodic groups. This data exists across multiple data types; formation energy is a continuous variable, periodic groups are ordinal, and the different types of stability (stable [blue], metastable against binaries [green], and metastable against elements [red]) are nominal.

To extend the single-feature clustermap algorithm to multiple data dimensions with mixed data-types, we used Gower's metric.[19,20] For each variable type, we choose a distance metric that is best suited to the data-type, and is scaled to fall between 0 and 1. For formation energy, which is continuous, we use the Euclidean distance. For the type of chemical family, which is an ordinal data type, we use the Manhattan distance on columns as defined in the left-step Periodic table,[21] and for the stability-type, which is a nominal data type, we use the Dice metric.[22] The Gower distance, *G*, is a linear combination of the various distance matrices,

$$G_{M_1-M_2} = \sum_i W_i \cdot d^i_{\mathrm{M_1\text{-}M_2}}$$

Where *W* governs the weights of the various contributions. In our work, we emphasized the clustering features in the following priority: formation energy of the stable compounds, formation energy of all compounds, chemical families, and then proximity of similar 'stability types' (as indicated on the map as blue/green/red), as well as the white spaces (where the DMSP did not identify a reasonable compound substitution).

A custom implementation of Gower's metric was coded within the *scipy* framework. Using the multi-feature distance matrix between the various elements, we calculate the linkages using the 'average' method (UPGMA algorithm) and construct a dendrogram with optimal leaf ordering.[23] We manually re-arranged the leaves of the dendrogram so as to order the elemental axes from left to right predominantly by Stable (Blue) → Metastable w.r.t binaries (Green) → Metastable w.r.t elements (Red). The resulting 1-D ordering of elements is set as the axis for the multi-feature clustermap.

## S.I.4 Ternary Nitride Discovery Histogram

Listed below are the years that each ternary $M_1$-$M_2$-N nitride space was discovered, as determined by the date of the publication for the first ternary nitride in that $M_1$-$M_2$-N system, as listed on the Inorganic Crystal Structure Database, shown in Figure 2d. Systems with stars are computed in DFT to be metastable (above the convex hull), corresponding to green or red squares in the ternary nitrides map.

1927 : C-Ca C-Ti*
1935 : Al-C
1938 : C-Na
1940 : C-K
1941 : C-Zn
1942 : C-Fe* C-Li
1945 : Au-C* C-Cd
1946 : Li-Mg Li-Zn
1948 : Ga-Li Al-Li
1949 : Li-Ni Cu-Li Co-Li
1950 : Nb-Ti* Ti-Zr* Nb-Zr* Nb-V* Ti-V*
1953 : Ge-Li Li-Si Li-Ti
1954 : Co-Mo Ni-Ti* Ni-Ta* Mo-Ni Co-Ti* Co-Ta* Mn-Ta* Ta-V* Fe-Ta*
1955 : Fe-Ni* Fe-Pt
1959 : Mn-Ni* Cr-Mn* Li-Mn Li-V Fe-Mn*
1960 : Hf-V* Hf-Zr* Hf-Ti* Hf-Ta* B-Zr* Fe-Pd
1961 : Cr-Li
1962 : Mn-Sn* In-Mn*
1963 : Al-Ti*
1964 : Hf-Zn Ag-Mn* Ti-Zn Nb-Zn Zn-Zr Ga-Ti In-Ti In-Zr* V-Zn Cr-Ga Ga-V* Al-Nb*
Cu-Mn*
1965 : Au-V*
1966 : Re-Sr Pd-S*
1967 : Ni-Zr* Fe-Nb* Co-Zr* Cr-Ta* Re-Zr* Os-Zr* Mn-Nb* Cr-Nb* Co-V* Ir-Zr* Ni-V* Ru-Zr*
Pd-Zr* Co-Nb* Fe-Zr* Hf-Nb* Nb-Ni* Rh-Zr* Pt-Zr* V-Zr*
1968 : Ge-Mn* Ga-Mn Fe-Ge*
1969 : Cr-Ge* Mn-Rh*
1970 : Ge-Mg Mg-Si Hf-Ni*
1971 : Mn-Pt* B-Ti* Cr-Ti* Mn-Si* Cr-Hf* Ge-V* Ge-Zn Mn-Zn Ca-Ge Li-Zr Cr-Zr*
1972 : Cr-Pd* Cr-Pt Mn-Pd* Cr-Sn* Cr-Rh* Ir-Mn* Cr-Ir*
1973 :
1974 : Nb-Ta* Ca-Ga
1975 : Cr-V* Ta-Ti* K-Zn Cs-Zn* Ta-Zr*
1977 : Sc-Ti* Mn-Ti* Cr-Sc* Sc-V* Mn-V*
1978 : Mo-Ta*
1979 : Mo-Nb*
1980 : Cd-K*
1981 : Cs-Sr*
1982 : Ba-Cs*
1983 : Rb-Zn*
1984 : Ca-Cs* Pb-S* Al-Hf* Ge-Na
1986 : B-Li Al-Zr*
1987 : In-Sr Ca-In
1988 : Sc-Ta* Co-Cs B-Nb*
1989 : Ca-Li Na-Ta Cs-Ta K-Ta Rb-Ta Li-Sr Nb-Sc*
1990 : Ca-Zn Ca-Cr B-Na Fe-Li Li-Mo Li-W Fe-Y* Ca-Fe Fe-Mo* Ca-Ni Ba-Ni
1991 : Ba-W Cu-Ta* Ga-Nb* Ba-Fe Li-Ta Ba-Mo B-Mg* Cu-Pd*

1992 : Na-W Sn-Zr* Li-Nb Bi-Ca Na-Nb Mo-Na Ca-Sb Fe-Sr Ca-Pb* Ca-V Ca-Sn Mg-Ta Ba-Na
1993 : Na-Si Mn-Sr Ca-Mn Ba-Mn Co-Y* Au-Ca Ba-Ta Cs-Nb
1994 : B-Sr Ba-Co Al-Sr Ba-Zr Ag-Ca B-Ca Fe-W Ba-Nb
1995 : Sr-Zn Ba-V Ca-Mg Ba-Cr Ni-Sr Ba-Si Sr-V Ba-Zn Ba-Ti Ba-Ga Ga-Sr Ca-Si Si-Sr Mn-W*
1996 : Sr-Zr Hf-Sr Ge-Sr Ba-Ge Cr-Sr Mn-Mo*
1997 : Cr-W* Cu-Sr
1998 : Ba-Hf Ba-Cu Sr-Ti Co-W* Nb-Sr Si-Zn Co-Sr Ni-W
1999 : Al-Ba Al-Ca
2000 : Na-Re Cs-Mn Rb-W Mo-Sr Cs-W Li-Re Cs-Mo
2001 : Mo-Pd* Mg-Sr
2002 : Au-Sr* Ba-Sr*
2003 : Sr-Ta In-Sc* S-Zr Li-Sc B-Ba
2004 : Ba-In Mo-Pt* Sb-Sr Ba-Bi Bi-Sr Ba-Sb
2005 : Na-Sn* Ba-Pb* Pb-Sr* Fe-Rh Ba-Sn* Sn-Sr*
2006 : Au-Cs* Ba-Li* Li-S* Au-Rb* Au-K* Ba-Rb*
2007 : Ca-Ti Al-B* Ca-Co Ca-Nb Sc-Zr*
2009 : Co-In* In-Ni* Fe-Ga*
2010 : Cs-Pd* Al-Ga* Li-Te* Li-Se*
2011 : Cd-Co Cd-Ni
2012 : Se-Zr Mg-Mo Al-Si*
2013 : Hf-S* Cu-Ni Ga-Mg*
2014 : Te-Y Fe-Se* Au-Cu*
2015 : Fe-Sn*

This Work Synthesized:
Mg-Hf, Mg-Ti, Mg-Zr, Mg-Nb, Zn-W, Mo-Zn, Sb-Zn

This Work Predicted (Stable Only):
Ir-Ba, Os-Ba, Re-Ba, Ru-Ba, Ru-Ca, K-Co, Na-Co, Co-Rb, Cr-K, Cr-Mg, Cr-Na, Cr-Rb, Cr-Zn, Y-C, Ge-K, Ge-Rb, Ca-Hf, Hf-Li, Na-Hf, Hf-Se, Li-In, Sr-Ir, K-B, K-Os, K-Pt, Re-K, Ru-K, K-Si, Bi-Li, Ir-Li, Os-Li, Ru-Li, Sb-Li, Y-Li, Mg-Zn, Mn-Co, K-Mn, Mg-Mn, Na-Mn, Mn-Rb, Mo-Ca, Mo-K, Mo-Rb, Y-Mo, Na-Ir, Na-Ni, Na-Os, Na-Ru, Na-Zn, Nb-K, Nb-Rb, Ca-Os, Cs-Os, Os-Rb, Sr-Os, Re-Ca, Cs-Re, Re-Mg, Re-Rb, Re-Zn, Cs-Ru, Ru-Rb, Ru-Sr, Sb-Mg, , Sc-C, Y-Se, Y-Si, Sn-Li, Mg-Sn, Zn-Sn, Ca-Ta, Zn-Ta, Cs-V, K-V, Mg-V, Na-V, Rb-V, Ca-W, K-W, Mg-W, Sr-W, Y-W, Ca-Zr, Na-Zr

## S.I.5 Deepest Hull Binary Nitrides

Formation energy of the 'deepest-hull' binary nitride, as computed in DFT-PBE and using reference data from the MaterialsProject database. In systems where a binary nitride is unknown (for example, Au-N), the deepest-hull formation energy is computed from a binary nitride structure generated using the same DMSP employed in this work, computed previously by the authors in Reference 2.

We note that in the Sr-N and Ba-N spaces, the deepest-hull binaries are actually $SrN_2$ ($\Delta H_f$ = -0.751 eV/at) and $BaN_2$ ($\Delta H_f$ = -0.765 eV/at). However, these compounds are diazenides,[24] which result in electronic structures that we anticipate to be poor reference states for the solid-state bonding analyses. For this reason, we use $Sr_3N_2$ and $Ba_3N_2$ as the reference binaries for the Sr-N and Ba-N systems, which are constructed in DFT based on the $Ca_3N_2$ structure. We note that these $Sr_3N_2$ and $Ba_3N_2$ compounds have comparable formation enthalpies with $SrN_2$ and $BaN_2$, as shown on the table below.

Convex hulls for the binary nitride spaces can be viewed for free on the Materials Project website, using the 'Phase Diagram' app.

| Element | $\Delta H_f$ (eV/atom) | Compound |
|---|---|---|
| Ag | 0.57 | $Ag_3N$ |
| Al | -1.59 | AlN |
| Au | 0.79 | AuN |
| B | -1.47 | BN |
| Ba | -0.76 | $Ba_3N_2$ |
| Bi | 0.24 | BiN |
| C | 0.14 | $C_3N_4$ |
| Ca | -0.95 | $Ca_3N_2$ |
| Cd | 0.39 | $Cd_3N_2$ |
| Co | -0.08 | CoN |
| Cr | -0.53 | CrN |
| Cs | 0.35 | $Cs_3N$ |
| Cu | 0.19 | $Cu_3N$ |
| Fe | -0.31 | FeN |
| Ga | -0.67 | GaN |
| Ge | -0.26 | $Ge_3N_4$ |
| Hf | -1.94 | $Hf_3N_4$ |
| In | -0.11 | InN |
| Ir | 0.16 | $IrN_2$ |
| K | 0.38 | $K_3N$ |
| Li | -0.46 | $Li_3N$ |
| Mg | -0.91 | $Mg_3N_2$ |
| Mn | -0.41 | MnN |
| Mo | -0.55 | MoN |
| Na | 0.26 | $Na_3N$ |

| Element | $\Delta H_f$ (eV/atom) | Compound |
|---|---|---|
| Nb | -1.25 | $Nb_2N$ |
| Ni | -0.06 | $Ni_3N$ |
| Os | 0.05 | $OsN_2$ |
| Pb | 0.33 | $Pb_3N_2$ |
| Pd | 0.23 | $PdN_2$ |
| Pt | 0.19 | $PtN_2$ |
| Rb | 0.48 | $Rb_3N$ |
| Re | -0.13 | $Re_3N$ |
| Rh | 0.11 | $RhN_2$ |
| Ru | 0.11 | $RuN_2$ |
| S | 0.35 | SN |
| Sb | 0.18 | SbN |
| Sc | -2.11 | ScN |
| Se | 0.63 | SeN |
| Si | -1.31 | $Si_3N_4$ |
| Sn | 0.00 | $Sn_3N_4$ |
| Sr | -0.75 | $Sr_3N_2$ |
| Ta | -1.37 | TaN |
| Te | 0.68 | $Te_3N_4$ |
| Ti | -1.91 | TiN |
| V | -1.16 | VN |
| W | -0.46 | WN |
| Y | -1.88 | YN |
| Zn | -0.03 | $Zn_3N_2$ |
| Zr | -1.87 | ZrN |

## S.I.6. Extended Metallicity/Ionicity/Covalency Figures

Figure S.I.6.1 van-Arkel Triangle for stable Alkali-Metal-Nitride Ternaries.

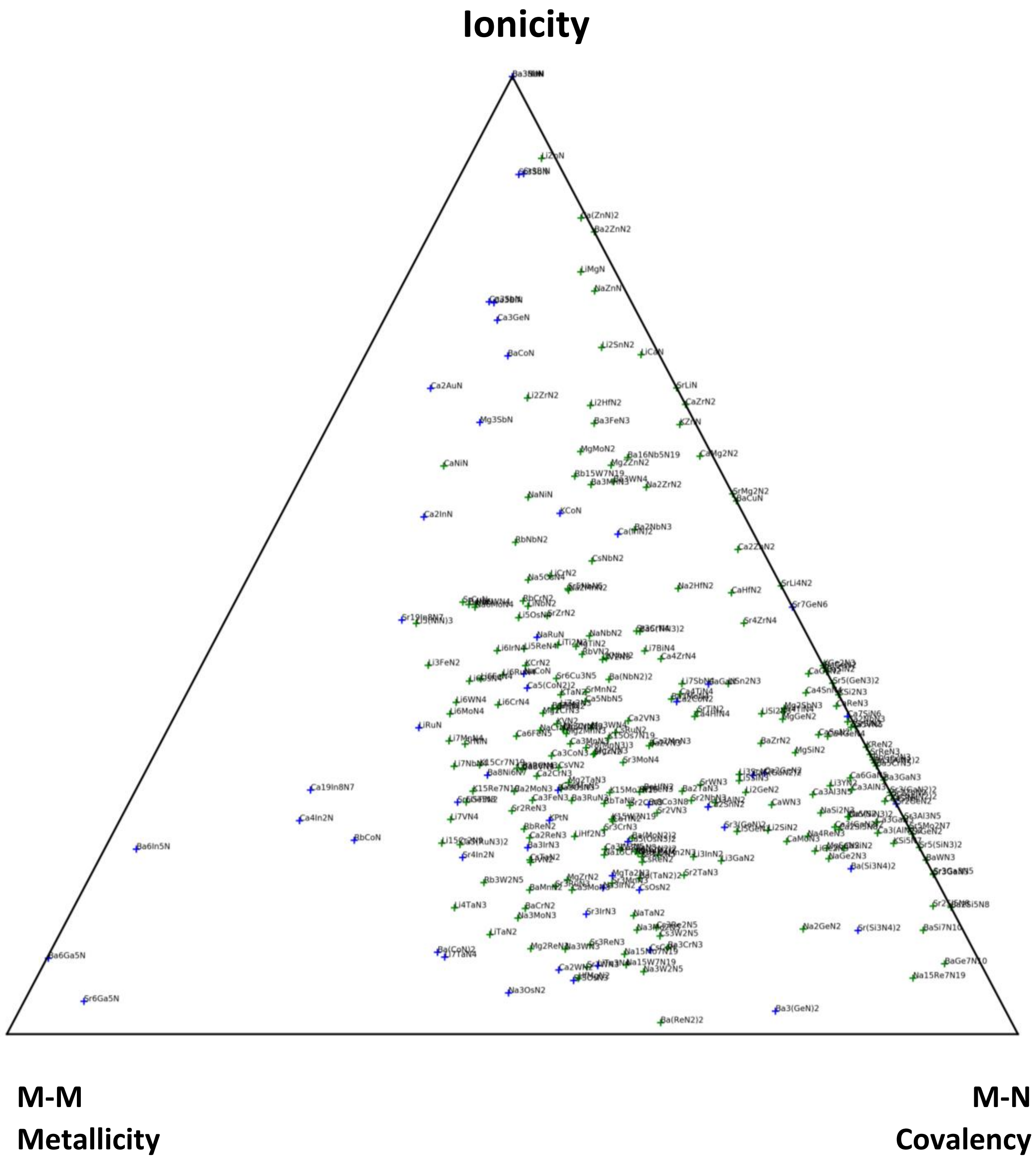

Figure S.I.6.2 van-Arkel triangle, stable non-alkali Metal-Metal-Nitride Ternaries.

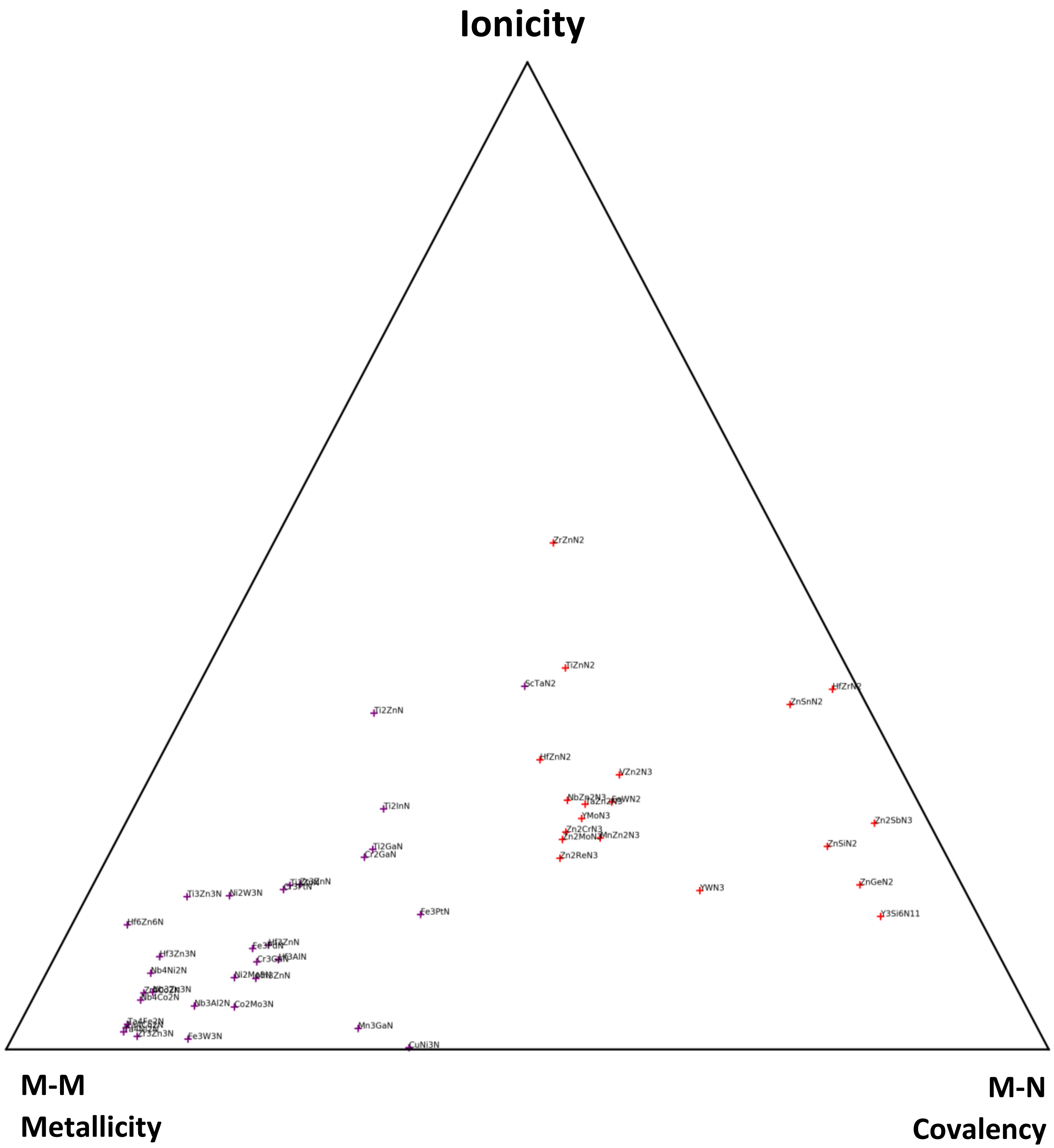

**Figure S.I.6.3** Scatterplot for change in ionicity of the more electronegative metal, *B*, and the nitrogen anion, *N*, with respect to the deepest-hull binaries, for stable Alkali-Metal-Nitride ternaries.

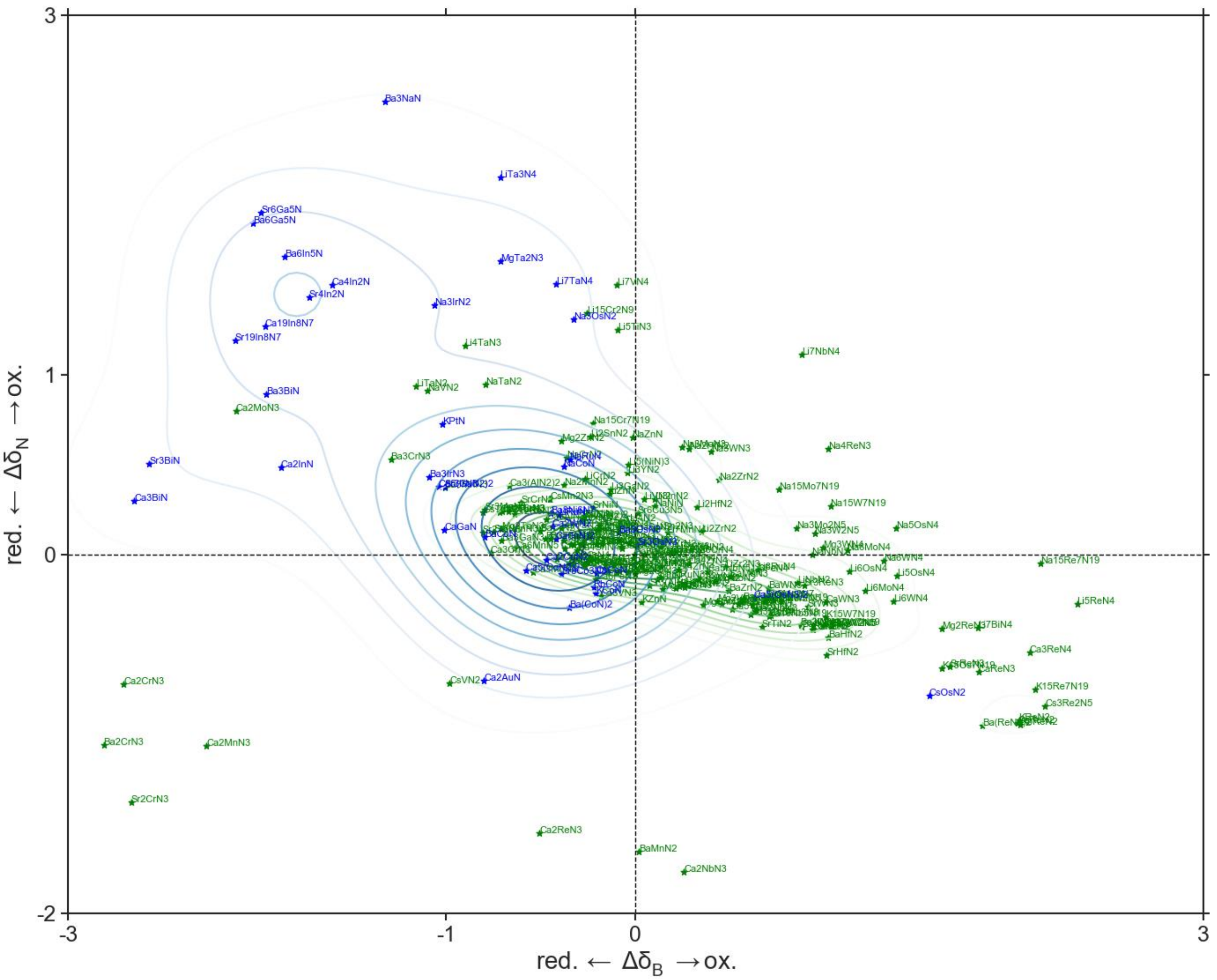

**Figure S.I.6.4** Scatterplot for change in ionicity of the more electronegative metal, $B$, and the nitrogen anion, $N$, with respect to the deepest-hull binaries. Shown for stable Alkali-Metal-Nitride ternaries, magnified on the -0.5 < ΔδN < 0.5, and -1.0 < $\Delta\delta_B$ < 1.5 ranges

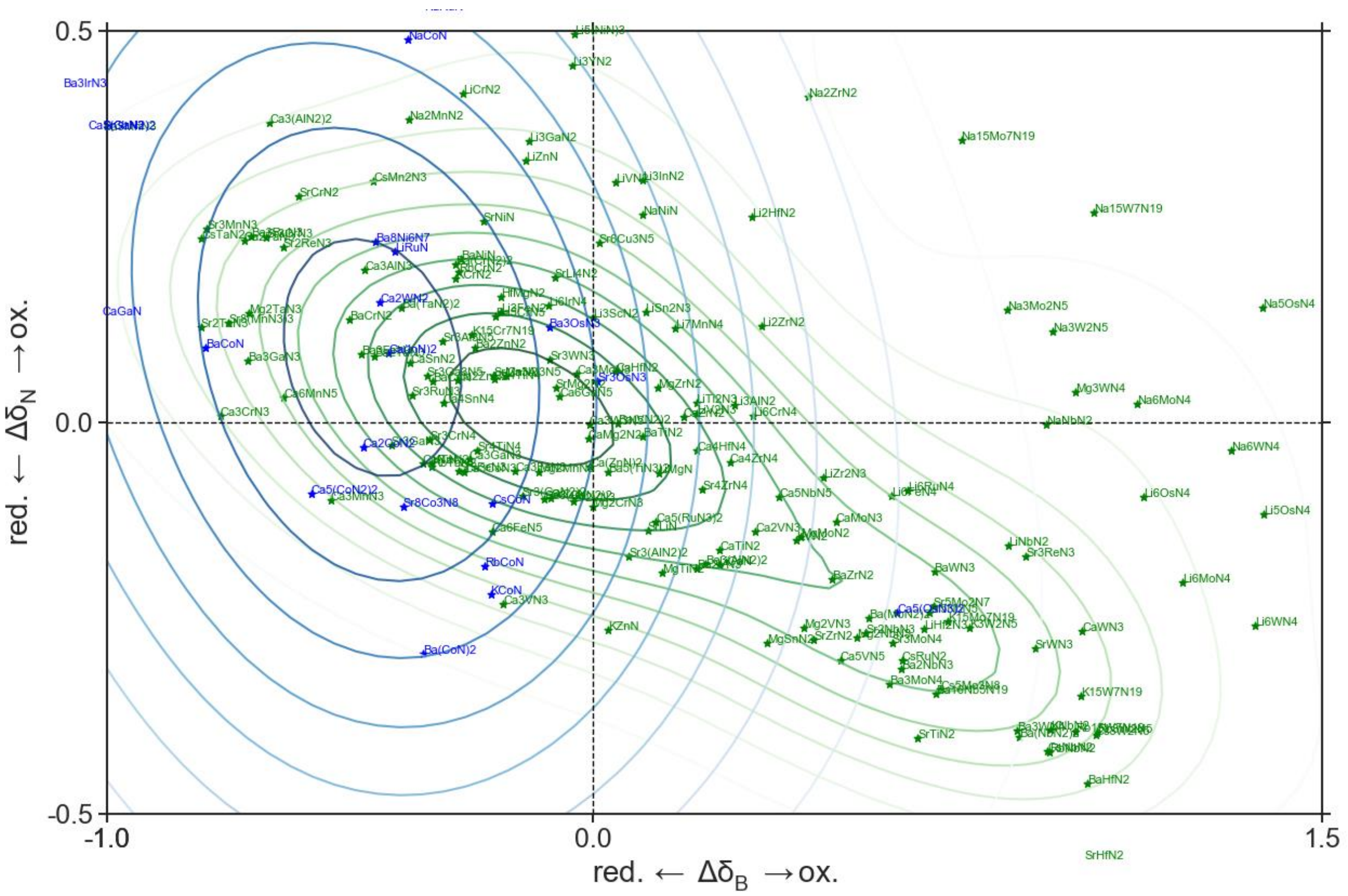

**Figure S.I.6.5** Scatterplot for change in ionicity of the more electronegative metal, *B*, and the nitrogen anion, *N*, with respect to the deepest-hull binaries.

Stable non-alkali Metal-Metal-Nitride Ternaries

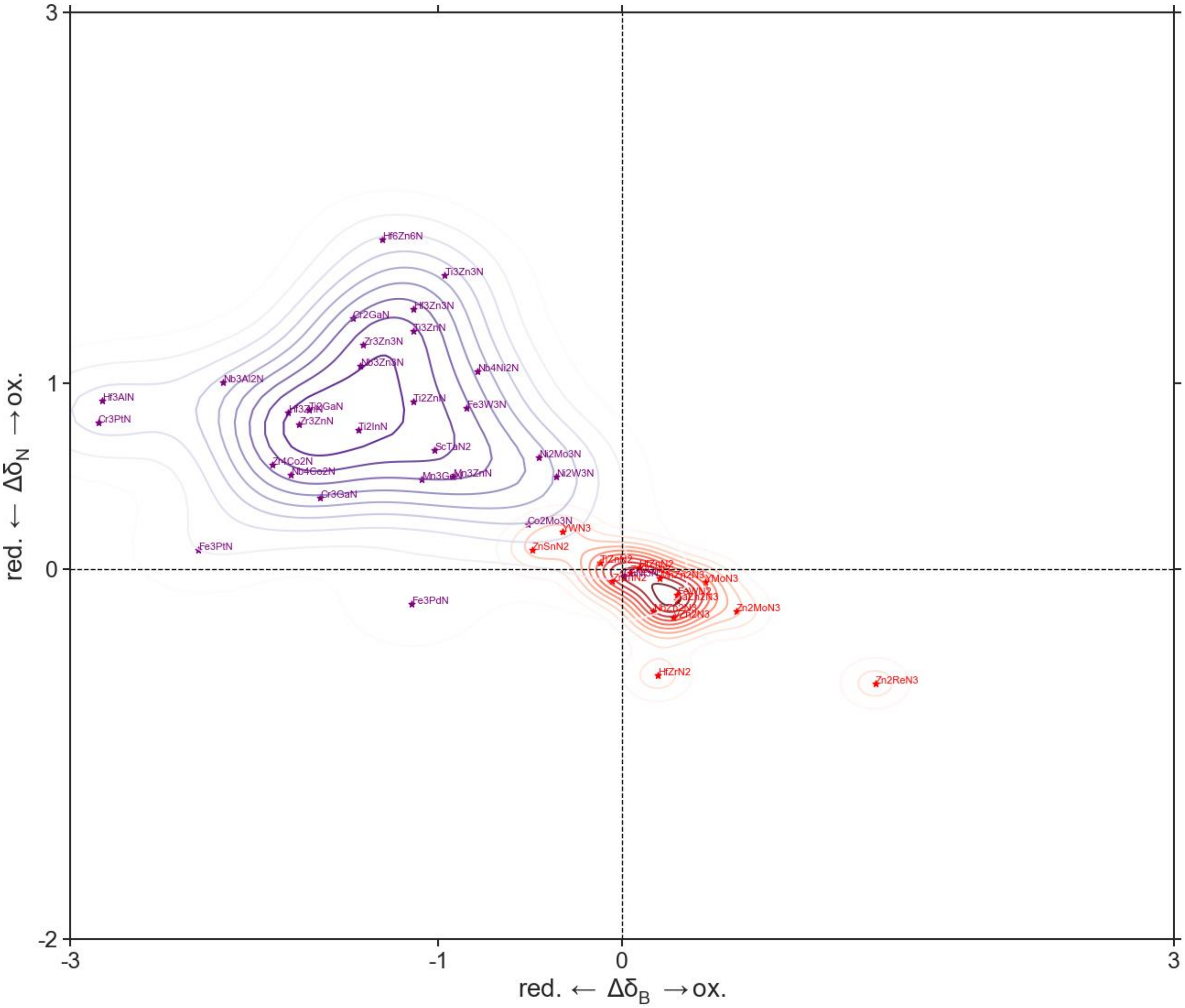

## S.I.7. *Reductive Effect*-stabilized structures

Formation of nitrogen-poor $Co_2Mo_3N$ and $Sr_3Ge_2N_2$ from the corresponding 'deepest-hull' binary nitrides proceeds by the following reactions:

$$2\ CoN + 3\ MoN \rightarrow Co_2Mo_3N + 2\ N_2$$

$$3\ Sr_3N_2 + 2\ Ge_3N_4 \rightarrow 3\ Sr_3Ge_2N_2 + 4\ N_2$$

In both systems, the metals are highly reduced compared to their corresponding binary nitrides, and there are strong metal-metal bonds in the ternary nitrides. As computed from the $\Delta\delta_N$ in **Figure 4.b**, the electrons in these metal-metal bonds can be explained by oxidation of nitrogen (with respect to the corresponding binary nitrides).

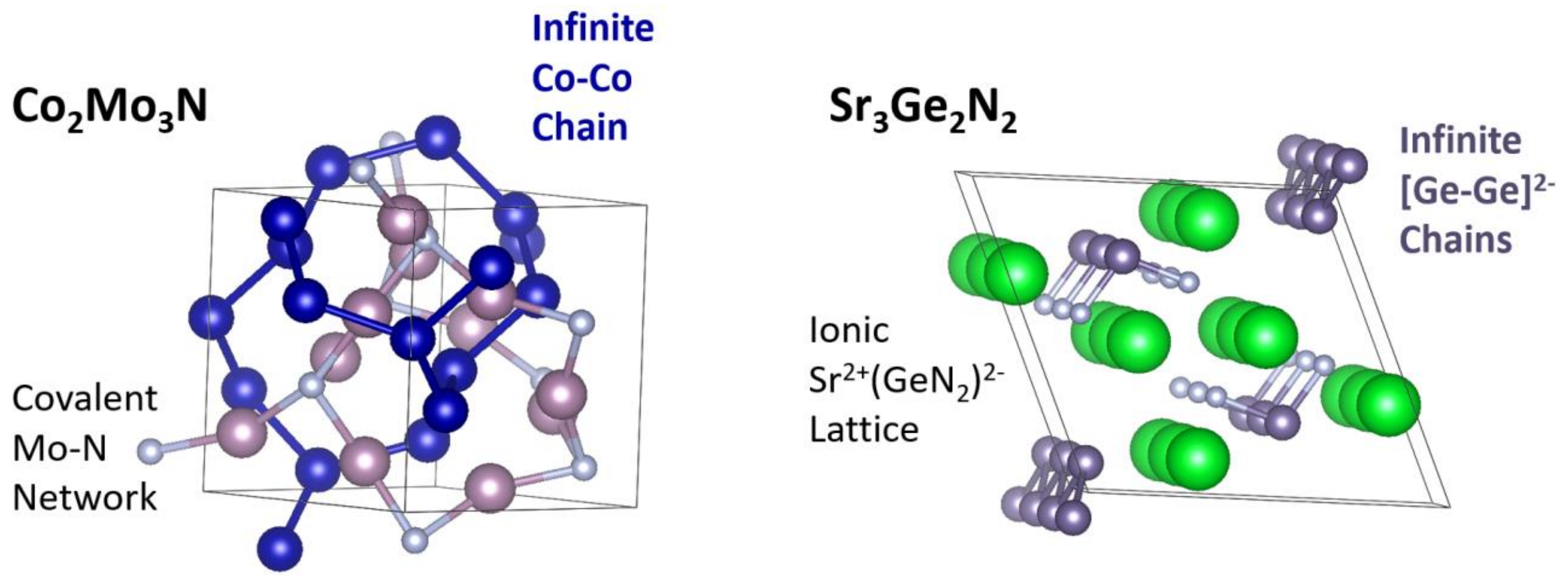

Supplemental Information References

[23] Bar-Joseph, Ziv, David K. Gifford, and Tommi S. Jaakkola. "Fast optimal leaf ordering for hierarchical clustering." *Bioinformatics* 17.suppl_1 (2001): S22-S29.

[24] Schneider, Sebastian B., Rainer Frankovsky, and Wolfgang Schnick. "Synthesis of alkaline earth diazenides MAEN2 (MAE= Ca, Sr, Ba) by controlled thermal decomposition of azides under high pressure." *Inorganic chemistry* 51.4 (2012): 2366-2373.